\documentclass[fleqn,usenatbib]{mnras}

\usepackage[T1]{fontenc}
\usepackage{ae,aecompl}
\usepackage{adjustbox}

\usepackage{epsf,palatino,url,wrapfig,array,setspace,amsmath,amssymb,fancyhdr,multirow,lscape,appendix,rotating}
\usepackage{graphics,epsfig,txfonts}
\usepackage{graphicx}

\usepackage{longtable}
\usepackage{amssymb}
\usepackage{xcolor}
\usepackage{color}
\usepackage{lipsum}
\usepackage{textcomp}
\usepackage{threeparttable,booktabs}
\usepackage{enumerate}
\usepackage{placeins}
\usepackage{float}
\usepackage[normalem]{ulem}
\usepackage{soul}
\usepackage{blindtext}
\usepackage{bm}

\usepackage{lipsum}  

\usepackage{tabularx}

\usepackage{tikz,hyperref}
\usepackage{academicons}
 \newcommand{\swift}{{\it Swift}\xspace}

\newcommand{\cxo}{\hbox{\it Chandra}\xspace}

\graphicspath{{./}{figures/}{plots/}}
 
\title[The rings of GRB~221009A]{Dust-scattering rings of GRB~221009A as seen by the Neil Gehrels \textit{Swift} satellite: can we count them all?}

\author[Vasilopoulos et al.]
{Georgios Vasilopoulos,$^1$\thanks{E-mail: georgios.vasilopoulos@astro.unistra.fr} 
Despina Karavola,$^2$ 
Stamatios I. Stathopoulos,$^2$ 
\newauthor{Maria Petropoulou$^{2,3}$}
\\
$^{1}$ Universit\'e de Strasbourg, CNRS, Observatoire astronomique de Strasbourg, UMR 7550, F-67000 Strasbourg, France \\
$^2$ Department of Physics, National and Kapodistrian University of Athens, University Campus Zografos, GR 15783, Athens, Greece \\
$^3$ Institute of Accelerating Systems \& Applications, University Campus Zografos, GR 15783, Athens, Greece
}

\date{Accepted XXX. Received YYY; in original form ZZZ}

\pubyear{2023}

\begin{document}
\label{firstpage}
\pagerange{\pageref{firstpage}--\pageref{lastpage}}
\maketitle

\begin{abstract} 
We present the first results for the dust-scattering rings of GRB 221009A, coined as the GRB of the century, as observed by the Neil Gehrels \swift satellite. We perform analysis of both time resolved observations and stacked data. The former approach enable us to study the expansion of the most prominent rings, associate their origin with the prompt X-ray emission of the GRB and determine the location of the dust layers. The stacked radial profiles increase the signal-to-noise ratio of the data and allows detection of fainter and overlapping peaks in the angular profile. We find a total of 16 dust concentrations (with hints of even more) that span about 15 kpc in depth and could be responsible for the highly structured X-ray angular profiles. By comparing the relative scattered fluxes of the five most prominent rings we show that the layer with the largest amount of dust is located at about 0.44 kpc away from us. We finally compare the location of the dust layers with results from experiments that study the 3D structure of our Galaxy via extinction or CO radio observations, and highlight the complementarity of dust X-ray tomography to these approaches.  
\end{abstract}




\section{Introduction}
 Gamma-Ray Bursts (GRBs) are the most energetic transient phenomena in the Universe. The prompt phase of the burst consists of intense gamma-ray flashes, and it can last up to hundreds of seconds in the case of long-duration events. While the exact mechanism for the production of the prompt gamma-ray spectrum is still under debate, it is commonly accepted that the prompt emission is produced within a relativistic collimated plasma outflow launched by the rotating central engine \citep[for a review see][]{2015PhR...561....1K}. As the plasma propagates in the interstellar medium (ISM) it sweeps up material, causing its gradual deceleration on timescales much longer than the prompt phase duration. This long-lasting emission, which is known as the afterglow, is observed over a wide range of energies (typically from X-rays to radio waves) and it is thought to be produced by synchrotron radiation of relativistic electrons accelerated at the external shock wave \citep{RM1992,CD1999}. Inverse Compton scattering of low-energy photons by relativistic electrons is typically put forward to explain the recent very high-energy ($E>100$~GeV) photons detections from a handful of GRB afterglows \citep[for a review see][]{2022Galax..10...66M}.

A very bright GRB was observed on October 9, 2022 by various instruments, including the \textit{Fermi} Gamma-ray Burst Monitor (GBM) and the Large Area Telescope (LAT) \citep{GRB221009A_GBM_GCN32642, GRB221009A_LAT_GCN32658}. The Burst Alert Telescope (BAT) of the Neil Gehrels \swift satellite detected a hard X-ray transient at $T_{\rm BAT}=59861.59$~MJD, i.e. about an hour later than GBM \citep{2022ATel15650....1D}.  
Overall, the prompt emission of GRB~221009A lasts about 330~s \citep{GRB221009A_GBM_GCN32642}. Preliminary gamma-ray light curves from KONUS-Wind  \citep{GRB221009A_KONUS_GCN32668} and AGILE \citep{GRB221009A_AGILE_GCN32650} show a precursor followed by two bright pulses (covering a period of about 100~s), and a fainter pulse starting at $\sim200$~s after the end of the bright episode. 
Observations of the afterglow with X-shooter at ESO's UT3 of the Very Large Telescope led to the determination of the burst's redshift $z=0.151$~\citep{2022GCN.32648....1D}. 
Moreover, according to \citet{2022GCN.32648....1D} multiple spectral features caused by the ISM of the Milky Way were detected, suggesting a large column density of Galactic material along our line of sight. 
The extreme brightness of this event complicates detailed spectral analysis with instruments like \textit{Fermi}-GBM and KONUS-Wind due to pile-up effects. Nonetheless, \citet{GRB221009A_KONUS_GCN32668} estimate the isotropic gamma-ray energy to be $E_{\rm iso}\sim 2\times 10^{54}$~erg using the GBM fluence reported by \citet{GRB221009A_GBM_GCN32642}. 
The combination of the proximity to us and the large energy output make this burst an extraordinary event~\citep[for comparison see Fig.~18 in][]{fermilat2nd}.
X-ray imaging of the afterglow with \textit{Swift}-XRT captured several bright rings around the burst's position \citep{2022ATel15661....1T}. 
These are formed by scattering of the X-ray burst emission by dust layers in our Galaxy in the direction of the source \citep[for a recent review on dust scattering and absorption, see][]{2022arXiv220905261C}. 

Dust scattering rings and halos have been used to study the ISM in the direction of bright X-ray transients with modern observatories \citep[e.g.][]{2015ApJ...806..265H,2016MNRAS.455.4426V,2016ApJ...825...15H,2016MNRAS.462.1847B,2017MNRAS.468.2532J,2018MNRAS.477.3480J,2019ApJ...875..157J,2021A&A...647A...7L}.
While this is not the first time that dust scattered rings were observed from a GRB \citep[see e.g.][and references therein]{Klose1994, Vaughan2004, Vianello2007}, the location of GRB~221009A on the sky ($l=52.96^{\rm o},b=4.32^{\rm o}$ in Galactic coordinates) and its large inferred isotropic gamma-ray energy offer a unique opportunity to study the Galactic dust via analysis of the ring structures. Here, we analyze publicly available data of \textit{Swift}-XRT obtained within a few days after the GRB trigger. Our goal is to determine the location of dust layers in the line of sight to the burst by studying the temporal evolution of the dust scattered rings. 

This paper is structured as follows. In Sec.~\ref{sec:model} we outline the geometrical model used for the description of the X-ray dust rings. In Sec.~\ref{sec:data} we present the data used for the construction of the angular X-ray surface brightness profiles, and describe the methods applied to the modelling of these profiles. We present our distance measurements in Sec.~\ref{sec:results}. We continue with a comparison of our results to those obtained from other probes of the dust content in the Galaxy,  and with a discussion on dust grain properties in Sec.~\ref{sec:discussion}. We finally conclude in Sec.~\ref{sec:summary} with a summary of our main findings.  

\section{Modelling of X-ray rings}\label{sec:model}
Dust is ubiquitous in the interstellar space but the largest dust concentrations (dust layers) are found inside dense cold molecular clouds. X-rays can be preferentially scattered or absorbed (depending on their energy) by interstellar dust grains. In this work we are interested in the geometrical study of the ring structures formed by dust scattering. Therefore we limit our analysis to photon energies $E \ge 1$~keV. We also neglect multiple X-ray scatterings by dust. 

\begin{figure}
\includegraphics[width=0.48\textwidth]{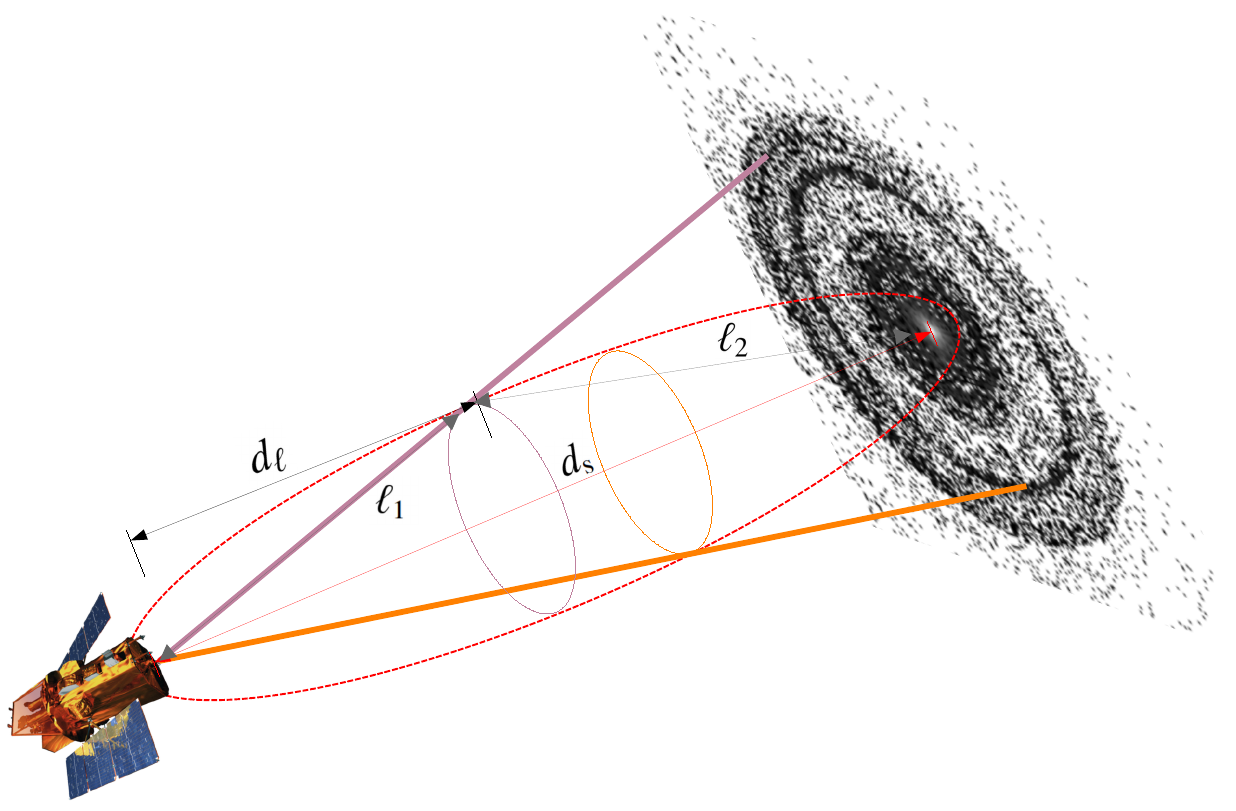} 
\caption{Schematic illustration (not in scale) of X-ray scattering by dust concentrated in layers located at different distances $d_{\ell}$ from the satellite. X-ray photons emitted by the GRB, which is located at a distance $d_s \gg d_\ell$, travel a distance $\ell_2$ before changing their direction due to scattering by dust at $d_{\ell}$. Then, scattered photons travel a distance $\ell_1$ before reaching the detector. The scattering of X-ray photons by different dust layers that are observed with the same time delay with respect to the burst defines an ellipsoid (red dotted line) with the satellite and the source as its two focal points. The projected image is a smoothed version of the XRT data analysed in this work.}
\label{fig:sketch}
\end{figure} 

 The geometrical principles of X-ray scattering by dust layers are illustrated in Fig.~\ref{fig:sketch}. We consider an X-ray transient occurring at time $t_{\rm b}$ and at a distance $d_{\rm s}$. X-ray photons can be scattered in small angles by an intervening dust layer at distance $d_\ell = x d_{\rm s}$, where  $x \ll 1$ for an extragalactic transient (e.g. $x=10^{-5}$ for a transient at 300~Mpc and a dust layer at 3~kpc from us). The scattered photons will be observed with a time delay $\Delta t$ with respect to the X-ray transient because of their longer path lengths,
\begin{equation}
    \Delta t =  \frac{\ell_1 + \ell_2}{c} - \frac{d_s}{c}
    \label{eq:delay}
\end{equation}
where $\ell_1 = d_\ell/\cos\theta = x d_{\rm }/\cos\theta$, $\ell_2 = \sqrt{(d_\ell \tan\theta)^2 + (d_{\rm s}-d_\ell)^2} = d_{\rm s} \sqrt{(x \tan\theta)^2 + (1-x)^2}$, and $\theta$ is the angular size of the ring (corresponding to the ring radius). For small angles ($\theta \ll 1$) the time delay can be approximated (up to second order in $\theta$) by the following expression
\begin{equation}
    \Delta t \approx \frac{d_{\rm s}}{2c} \frac{x}{1-x}  \theta^2 \approx \frac{d_{\ell}}{2c}\theta^2. 
\end{equation}
Photons scattered by the same dust layer but arriving with larger time delays will produce a ring of larger angular size. In other words, each ring produced by a single dust layer appears to expand with time. Using the equation above, and assuming $x\ll 1$, we find an expression for the time evolution of $\theta$
\begin{eqnarray}
\theta & \simeq  & 4.8~{\rm arcmin} \left(\frac{x}{10^{-6}}\right)^{-1/2} \left(\frac{d_{\rm s}}{100~\rm Mpc}\right)^{-1/2} \left(\frac{\Delta t}{10^4~\rm s}\right)^{1/2}  \nonumber  \\
&= &  4.8~{\rm arcmin} \left(\frac{d_{\ell}}{100~\rm pc}\right)^{-1/2} \left(\frac{\Delta t}{10^4~\rm s}\right)^{1/2}
\label{eq:theta}
\end{eqnarray}
The surface of equal time delays is an ellipsoid with the telescope and the X-ray source placed at the two focal points. Therefore, if multiple dust clouds intersect this surface will produce separate rings of different angular sizes by photons arriving to the observer with the same time delay (see Fig.~\ref{fig:sketch}). At any given time rings observed with smaller angular sizes are those produced by the more distant layers and vice versa.

Throughout the analysis we adopt a value of $z=0.151$ for the GRB redshift, which corresponds to a luminosity distance $726.5$~Mpc (or a light travel distance $d_{\rm s}=585.6$~Mpc) based on WMAP9 cosmological parameters \citep{wmap9}. Eqs.~(\ref{eq:delay})-(\ref{eq:theta}) neglect redshift corrections, since the dust scattering layers are located in the Galaxy \citep[see also][]{1966MNRAS.132..101R,Vaughan2004}.

\section{Data reduction and analysis}\label{sec:data}
We use data from the Neil
Gehrels \swift satellite X-ray telescope \citep[\swift-XRT,][]{2005SSRv..120..165B}. These were retrieved from the \swift \ science data centre\footnote{\url{http://www.swift.ac.uk/user_objects/}} and analyzed using standard procedures as outlined in \citet{2007A&A...469..379E,2009MNRAS.397.1177E}. 
We use five XRT observations performed between MJD 59862 -- 59866 with obs-id numbers 01126853004, 01126853005, 01126853006, 01126853008 and 01126853009. From the cleaned images we selected events (grade 0-12) with energies between 1 and 10 keV and barycentric corrected times. 

Our analysis relies on radial profiles of X-ray photons. Determination of the source's position in the image (i.e. the actual center of the rings) is therefore crucial. Another important effect is the quite rapid expansion of the rings; their angular diameter can evolve significantly on timescales of less than a day -- see Eq.~(\ref{eq:theta}). We thus split observations into groups of events obtained within a time window of less than 20~ks. We end up with 10 useful subsets of data. We perform source detection and localization in each subset of \swift-XRT data, and compute the respective exposure maps. Upon correcting each data subset with the appropriate exposure map, we compute radial profiles of X-ray surface brightness (in units of counts s$^{-1}$ arcmin$^{-2}$). 

\begin{figure*} 
\centering
\includegraphics[width=\textwidth, trim= 0 50 0 0 ]{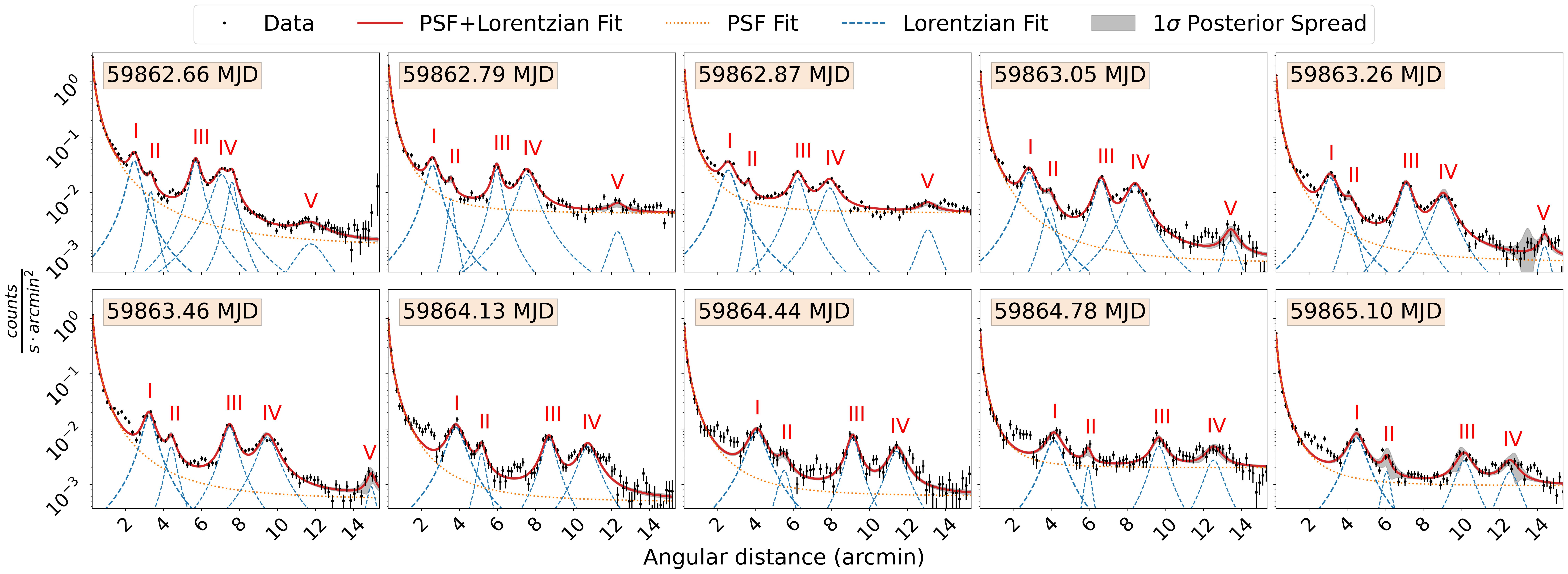}
\caption{Time evolution of the angular X-ray surface brightness profile constructed using X-ray photons with $E\ge1$~keV. For each observation the optimal model (solid red curve), and its decomposition into the various components (dashed blue and orange lines), is overplotted. The peaks of the most prominent (primary) rings identified in the observations are indicated with numbers in each panel. The fourth ring is fitted with two Lorentzians only in the first dataset, since these could not be securely identified in the following datasets. The grey shaded band in each panel indicates the 68 per cent confidence interval.
}\label{fig:radialP}
\end{figure*}

\subsection{Modelling of radial profiles}\label{sec:methods}
To model the radial profile of the X-ray surface brightness (in units of counts s$^{-1}$ arcmin$^{-2}$) we use the updated point-source function (PSF) for \swift-XRT\footnote{\url{https://www.swift.ac.uk/analysis/xrt/pileup.php}}, 
\begin{equation}
f_{\rm PSF}(r) = A \left [W e^{-\frac{r^2}{2 \, \sigma^2}}+ (1-W) \left( 1+ \left(\frac{r}{r_{\rm c}}\right)^2\right)^{-b}\right] + B
\label{eq:psf}
\end{equation}
where $W=0.075$,  $\sigma =7.42$~arcsec, $r_{\rm c}=3.72$~arcsec, and $b \sim 1.31$. In the fitting procedure we leave the power-law index $b$ free to vary and introduce an additional normalization parameter $A$ to account for possible pile up in the detector.  We also add a constant $B$ to account for possible contribution of the background. Each distinctive peak in the angular profile, which corresponds to a ring in the XRT image, is modelled with a Lorentzian function
\begin{equation}
f_{\rm L}(r) = \frac{a_{\rm L}}{\pi} \frac{c_{\rm L}}{(r-b_{\rm L})^2+c_{\rm L}^2}
\label{eq:lor}
\end{equation}
where $a_{\rm L}$ is the normalization, $b_{\rm L}$ is the position of the peak, and $2c_{\rm L}$ is the full width at half maximum. The final fitting function applied to the angular profiles is
\begin{equation}
f_{\rm tot}(r) = f_{\rm PSF}(r) + \sum_{i=1}^{n}f_{\rm L_{i}}(r)
\label{eq:model}
\end{equation}
where $n$ is the total number of peaks.

\begin{figure*}
\centering
\includegraphics[width=0.98\textwidth ]{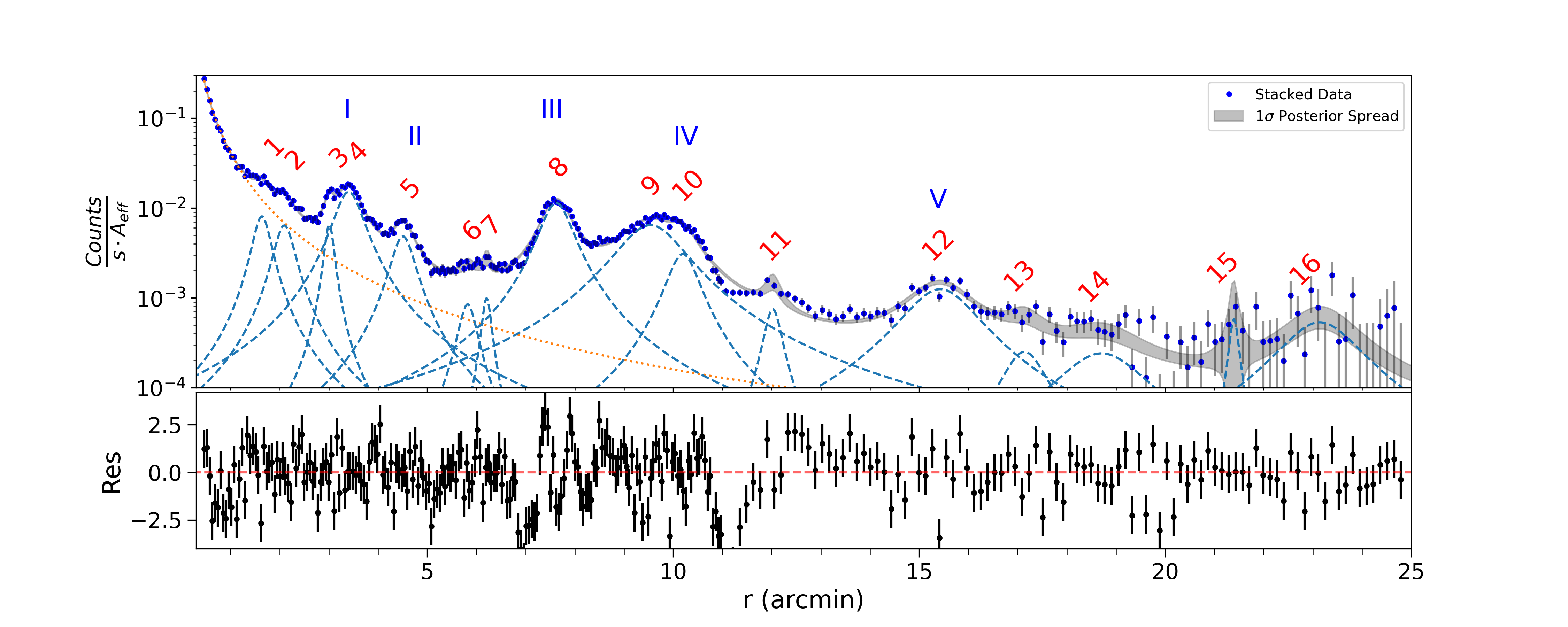}
\caption{\textit{Top panel:} Angular profile of the X-ray surface brightness using the stacked X-ray image shown in Fig.~\ref{fig:im_stack} at a reference time of 2 days since GBM trigger. The optimal model is decomposed into the PSF (dotted line) and multiple Lorentzians  (dashed lines) that are also indicated by Arabic numbers. Roman numbers are used for noting the peaks from the time-resolved angular profiles in Fig.~\ref{fig:radialP}. The grey shaded region indicates the 68 per cent confidence interval computed from the posterior samples. 
\textit{Bottom panel:} Plot of residuals computed as the ratio of the difference between the optimal model and the data to the error. }
\label{fig:stacked}
\end{figure*}
\subsection{Analysis of individual datasets}
To identify significant peaks in the radial profile distribution we use an iterative process. We start with a radial profile and smooth it with a Savitzky-Golay filter to eliminate noise \citep{1964AnaCh..36.1627S}. We then identify prominent maxima in the smoothened radial profile (in logarithm) above a certain threshold (i.e. 0.05 in dex) compared to local neighbouring values.
As our goal is to identify prominent peaks we are conservative on the choice of the threshold level. In other words, a lower threshold would lead to a few more peaks that would be consistent with noise. 

We then construct a model composed of the PSF and Lorentzian functions -- see Eq.~(\ref{eq:model}) -- centered at the locations of the identified peaks. We optimize the model to the data (without any smoothing) with a least-square algorithm. We then construct a residual plot with the values normalised over the data uncertainties. Structures in the residual plot can help us identify  secondary peaks. We repeat the procedure to search for secondary peaks in the data above a 3$\sigma$ level (i.e. 3 times above the errors of each point). This step is crucial since some peaks might be missed initially because they are either very close to other prominent peaks or their peak is hidden by the decay in intensity of the PSF profile, leaving only the side lobes visible. 

We then use the complete model (composed of the PSF and all peaks identified so far) and fit the profiles of each dataset once again using {\tt emcee} \citep{2013PASP..125..306F}, a python implementation of the Affine invariant Markov chain Monte Carlo (MCMC) ensemble sampler. This allows us to better estimate the uncertainties in model parameters and to explore possible degeneracies in this multi-parameter problem. 
 
The iterative procedure described above is applied only to the first dataset with the highest photon statistics. The optimal model is then used as an initial guess for the MCMC sampling of the next dataset. All parameters are sampled from uniform distributions in linear space, except for the background $B$ which is sampled from a uniform distribution in log-space. 
We produced a chain with 200 walkers that were propagated for 2500 steps each;
after testing we concluded that this is an optimal number of steps for the convergence of the walkers. We also discard the first 1000 steps of each chain as burn-in.

We present in Fig. \ref{fig:radialP} the angular profiles for 10 individual datasets with the MCMC fitting results overlaid, and list the optimal model parameters for the Lorentzians in Table~\ref{tab:optimal_params}. In the first angular profile we clearly identify 5 prominent peaks. The fourth ring can be described by two Lorentzian functions. However, we neglect this substructure since these two distinct components are not observed in the following datasets.  As the time progresses the rings are expected to grow apart thus allowing us to to see more structures in the angular profiles, i.e. secondary rings -- see e.g. the bump appearing in the lower panels of Fig.~\ref{fig:radialP} at smaller angular distances than the first ring. Meanwhile other rings, like the fifth one, can move outside the field of the CCD camera as they expand. It is also possible that some of the dust scattering rings disappear as their intensity faints or due to changes in the ISM properties as each snapshot maps dust scattering at different locations. The spread in the modelled angular profiles becomes larger around peaks at large angular distances where the statistical errors become larger (see e.g. last panel from the left in the top row of Fig.~\ref{fig:radialP}). This spread is also suggestive of the presence of substructure in the outer rings. A complementary stacking analysis of XRT data, which is presented in the next section, can help us search for such features in the combined angular profile.

\begin{figure}
\centering
\includegraphics[width=0.5\textwidth]{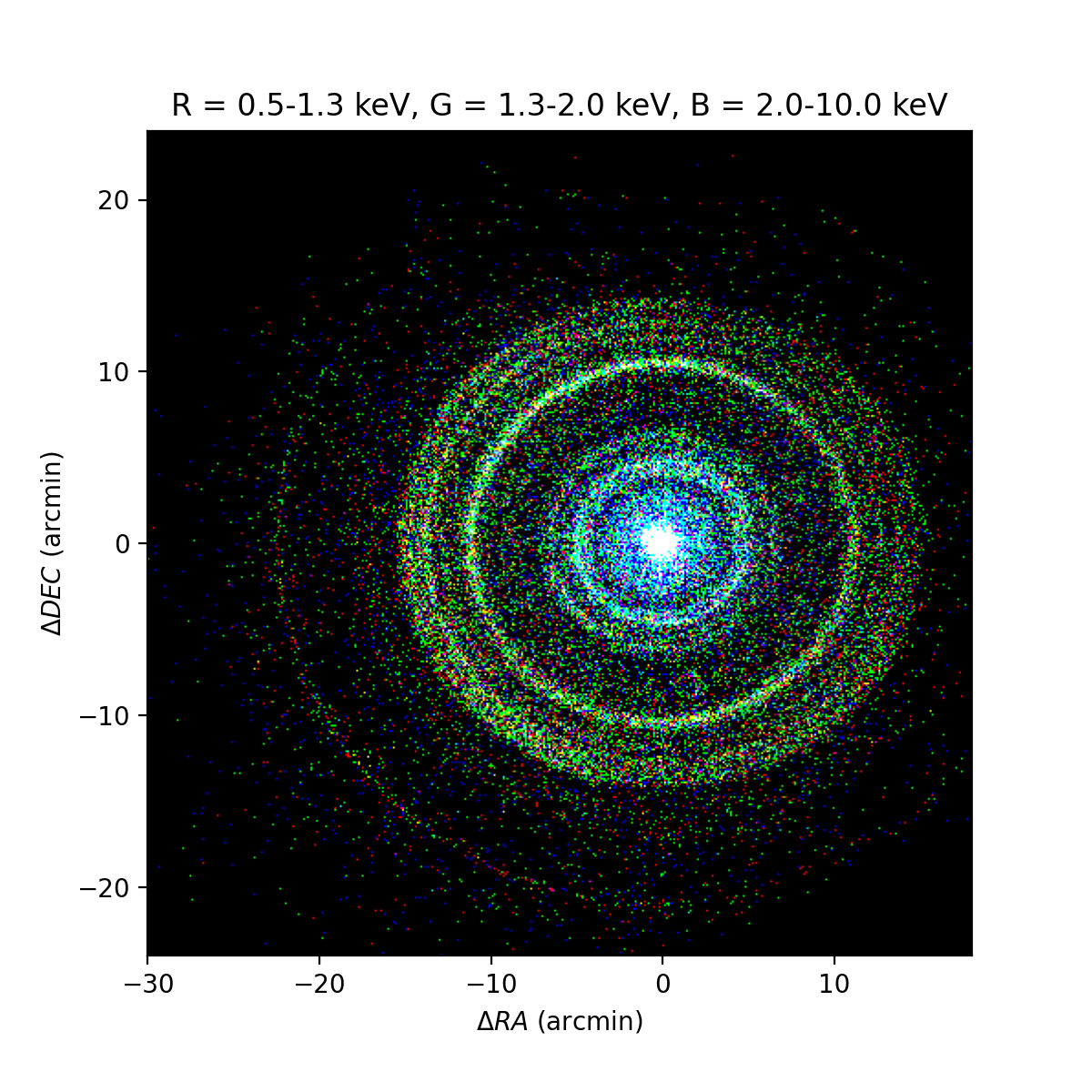}
\caption{
Stacked RGB image of all \swift-XRT observations used in our analysis for a reference time of 2~days post GBM trigger. The image is centered on the GRB location. The position of each photon has been re-scaled from the center of the image (0,0) using the ring expansion law $\theta \propto \sqrt{\Delta t}$  given in Eq.~(\ref{eq:theta}). We use the 0.5-10 keV energy range only for this image creation.
}
\label{fig:im_stack}
\end{figure}

\subsection{Stacking analysis of all data}
An alternative method to identify dust echoes is to stack all XRT images in order to increase the signal to noise. However, this is not as simple as adding the images because of the dynamic nature of the problem.
Assuming each and every photon above 1 keV  was scattered once in an intervening dust layer, we can scale its position on the image at an arbitrary time based on the expansion law of Eq.~(\ref{eq:theta}) and the time the photon was recorded. We define the position of each photon in the image using polar coordinates ($r$, $\phi$) centered at the location of the GRB. Using the time of arrival of each event we re-scale the $r$ coordinate to $r_{\rm rs} = r \, (\Delta{t_{\rm rs}}/\Delta{t_{\rm event}})^{1/2}$, where $\Delta{t_{\rm rs}}$ is the reference time for the re-scaled stacked image and $\Delta{t_{\rm event}}$ is the time delay between the detection of the photon and the burst. As an indicative example we select $\Delta{t_{\rm rs}}=2$~d and use the GBM trigger time as reference time for the GRB, i.e. 13:16:59.99~UT on 09 October 2022 or $T_{\rm GBM} =59861.55$~MJD~\citep{GRB221009A_GBM_GCN32642} -- the choice of this reference time will become clearer in the next section. 

The stacking procedure increases the signal to noise in the outer regions, thus enabling us to extend the radial profiles up to a radius of $\sim25$ arcmin, as illustrated in  Fig. \ref{fig:stacked}. We also use an adaptive binning for the stacked angular profile with denser sampling for the inner part (i.e. $\sim$4 versus $\sim$20~arcsec) for a clearer presentation. After correcting the stacked radial profiles using the individual exposure maps of each snapshot, we follow the same procedure described in the previous section to identify features that could be related to X-ray rings. The analysis of the stacked image, which is shown in Fig.~\ref{fig:im_stack}, leads to the identification of 16 Lorentzians (see dashed lines in Fig. \ref{fig:stacked}) that will be discussed further in the following section.
A model based on Eq.~\ref{eq:model} was fitted to the radial profiles with a similar procedure as the one described in the previous section, so all parameter quoted are based on the MCMC modelling.

\section{Localization of dust layers}\label{sec:results} 
We fit the temporal evolution of the angular radii of the five most prominent rings identified in individual XRT images (Fig.~\ref{fig:radialP}) using {\tt emcee} and the expansion law of Eq.~(\ref{eq:theta}). The statistical uncertainties of the Lorentzian centers (see Table~\ref{tab:optimal_params}) typically underestimate the uncertainty introduced by our model selection (e.g. PSF with 4 or 5 Lorentzians) and the poor knowledge of priors.  
Thus, when modelling the ring expansion, we add a term $\ln{f}$ to the likelihood function to account for the systematic scatter and noise not included in the statistical uncertainties of the estimated angular radii \citep[see similar application][]{2022MNRAS.tmp.3021K}, 
\begin{equation}
\ln{\mathcal{L}} =  -\frac{1}{2} \sum_{i}^{} \frac{({\rm model}-{\rm data})^2}{\sigma_{\rm tot, i}^2}+e^{\sigma_{\rm tot, i}^2}. 
\end{equation}
Here, the total variance is defined as 
\begin{equation}
\sigma_{\rm tot, \rm i}^2 = \sigma_{\rm i}^2 + e^{2\ln{f}}, 
\end{equation}
where $\sigma_{\rm i}$ are the errors of the Lorentzian centers $b_{\rm L, i}$. 

\begin{figure}
\centering
\includegraphics[width=0.47\textwidth]{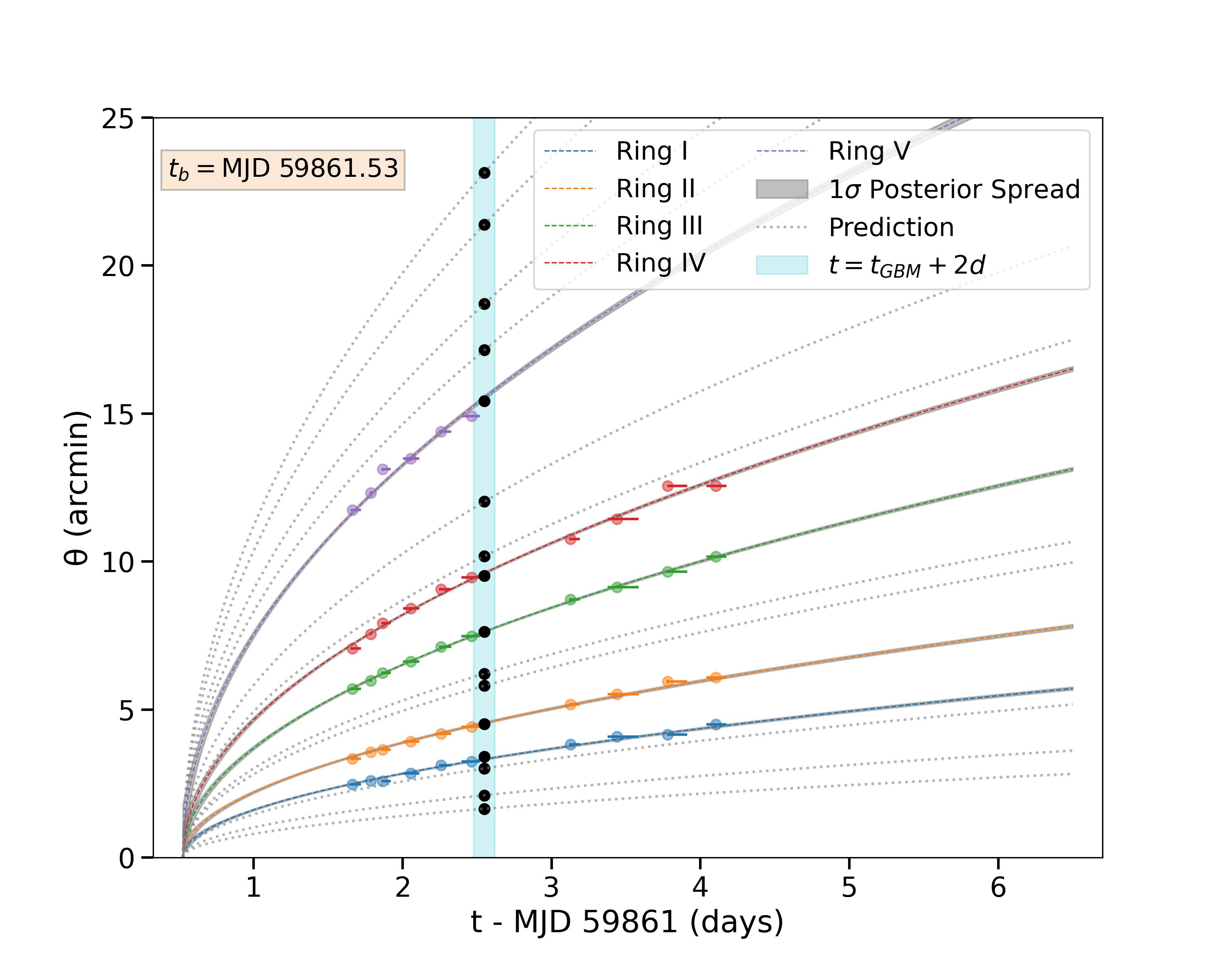}
\caption{Temporal evolution of the angular size of rings detected in individual \swift-XRT images (coloured symbols). 
The most probable model for each ring is overplotted with a dashed line of the same colour as the markers. The grey shaded regions indicate the 99.7 per cent confidence intervals. Black markers indicate the rings identified in the stacked image at a reference time of 2 days post GBM trigger. Dotted grey curves show the predicted expansion according to Eq.~(\ref{eq:theta}).}
\label{fig:expansion}
\end{figure}

Our optimal expansion model for each ring is shown in  Fig.~\ref{fig:expansion} (see coloured lines), the corner plot with the posterior distributions of all layers is presented in Fig.~\ref{fig:corner-2} and the dust layer distances are listed in Table~\ref{tab:params}. The derived time of the burst is $t_{\rm b}= {\rm MJD}~59861.53 \pm 0.02$, which is about one hour and a half earlier than the BAT trigger time $T_{\rm BAT} =59861.59$~MJD and consistent within errors with the GBM trigger time $T_{\rm GBM} =59861.55$~MJD~\citep{GRB221009A_GBM_GCN32642}. 
Therefore, the rings imaged by XRT are produced by X-rays emitted in the prompt phase of the GRB and scattered by dust in our Galaxy. This demonstrates that X-ray photons with energies down to 1 keV are produced during the prompt phase of GRB~221009A, even though they could not be detected by BAT and XRT simultaneously with GBM. Extension of the MeV gamma-ray spectrum to soft X-rays is a common prediction of radiative models, but the prompt X-ray fluence depends on the model details \citep[see, e.g.,][for lepto-hadronic radiative models of GRB~221009A]{2022arXiv221200766R}.

In regard to the stacked analysis we have demonstrated that by appropriate rescaling of the XRT images we can maintain the information of the peak locations and increase the signal to noise, enabling us to identify more structure in the data. For example several features that appear only in a few snapshots (see Fig.~\ref{fig:radialP}) are enhanced in the stacked profiles. In Fig.~\ref{fig:stacked} we can identify at least 8 prominent humps, with one of them  being clearly double peaked (composed of peaks \#3, \#4) and some of them being quite broad (i.e. \#9, \#10 and \#12). The angular sizes of all identified peaks and the distances of the corresponding dust scattering locations are summarized in Table~\ref{tab:params-stacked}. If we consider that the sizes of the rings are just a projection effect, we need to use the estimated distances in order to ascertain if two nearby rings may be associated with the same dust layer and appear as separate due to inhomogeneities in the dust distribution of a single cloud. In fact the four innermost rings that appear to overlap the most in the angular profile are those that are physically the most detached, since the relevant dust layers are located at distances of about 14.7 kpc, 9.07 kpc, 4.4 kpc and 3.4 kpc. Thus, they cannot be associated with the same production site.

The innermost peak of the stacked data is also seen in individual snapshots (see e.g. the last two panels in the bottom row of Fig.~\ref{fig:radialP}), but its structure does not remind that of an extended halo. To check if these innermost peaks follow the $\sqrt{\Delta} t$ expansion law, we performed an additional fit to the last 4 individual datasets by adding two more Lorentzian functions. However, the Lorentzian centers do not seem to follow the expansion law.
Given that our results are limited by the \swift/XRT angular resolution, the origin of these features should be revisited with follow-up analysis of \cxo data.

\renewcommand*{\arraystretch}{1.1} 
\begin{table}
\centering 
\caption{Dust layer distances obtained from fitting the expansion of the most prominent rings in individual images using Eq.~(\ref{eq:theta}).} 
\label{tab:params}
\begin{threeparttable}
\begin{tabular}{c c c}
\toprule
Ring  & $x$ & $d_\ell$~(kpc) \\
\hline 
I  & $(6.23 \pm 0.11)\times 10^{-6}$ &${3.65 \pm 0.06}$ \\
II  & ${(3.32 \pm 0.05)}\times 10^{-6}$ & ${1.95 \pm 0.03}$ \\
III  &  ${(1.18 \pm 0.02)}\times 10^{-6}$ & ${0.691 \pm 0.008}$ \\
IV  & ${(7.45 \pm 0.09)}\times 10^{-7}$ & ${0.436 \pm 0.005}$ \\
V  & ${(2.85 \pm 0.04)}\times 10^{-7}$ & ${0.167 \pm 0.003}$\\
\bottomrule
\end{tabular}
\begin{tablenotes}
\item {Note. --} The listed values and errors correspond to the median value and the 68 per cent range of the posterior distributions, respectively. 
\end{tablenotes}
\end{threeparttable}
\end{table}

\begin{table}
\centering 
\caption{Locations of dust scattering regions obtained from fitting all peaks in the angular profile of the stacked X-ray image.} 
\label{tab:params-stacked}
\begin{threeparttable}
\renewcommand{\arraystretch}{1.3}
\begin{tabular}{p{0.2\columnwidth}p{0.3\columnwidth}p{0.3\columnwidth}}
\toprule
Ring\tnote{*} & $\theta$ (arcmin) & $d_\ell$~(kpc)\\
\hline
1 & 1.644$_{- 0.03 }^{+ 0.011 }$& 14.7$_{- 0.2 }^{+ 0.5 }$\\ 
2 & 2.095 $_{- 0.02 }^{ 0.011 }$& 9.07$_{- 0.09 }^{+ 0.2 }$\\
3 (I) & 3.009$_{- 0.012 }^{+ 0.008 }$& 4.40$_{- 0.02}^{+ 0.03}$\\
4 (I) & 3.402$_{- 0.013 }^{+ 0.005 }$& 3.440$_{- 0.011 }^{+ 0.03 }$\\ 
5 (II) & 4.505$_{- 0.011 }^{+ 0.009 }$& 1.961$_{- 0.008 }^{+ 0.009 }$\\ 
6 & 5.804$_{- 0.03 }^{+ 0.020 }$& 1.182$_{- 0.007 }^{+ 0.012 }$\\ 
7 & 6.209$_{- 0.009 }^{+ 0.020 }$& 1.033$_{- 0.007}^{+ 0.003 }$\\ 
8 (III) & 7.624$_{- 0.008 }^{+ 0.008 }$& 0.6849$_{- 0.0014 }^{+ 0.0014 }$\\ 
9 (IV) & 9.538$_{- 0.03 }^{+ 0.011 }$& 0.4376$_{-0.0011 }^{+ 0.002 }$\\ 
10 (IV) & 10.19$_{- 0.03 }^{+ 0.02 }$& 0.3835$_{-0.0008 }^{+ 0.002}$\\ 
11 & 12.02$_{- 0.02 }^{+ 0.03 }$& 0.2753$_{- 0.0013}^{+ 0.0011 }$\\
12 (V) & 15.43$_{- 0.03 }^{+ 0.02 }$& 0.1673$_{- 0.0004}^{+ 0.0006}$\\ 
13 & 17.14$_{- 0.02 }^{+ 0.03 }$& 0.1356$_{- 0.0004 }^{+ 0.0003 }$\\
14 & 18.71$_{- 0.02 }^{+ 0.02 }$& 0.1138$_{- 0.0002 }^{+ 0.0002 }$\\ 
15 & 21.41$_{- 0.02 }^{+ 0.04 }$& 0.0869$_{- 0.0003 }^{+ 0.00014 }$\\ 
16 & 23.14$_{- 0.02 }^{+ 0.02 }$& 0.07434$_{- 0.00014 }^{+ 0.00006 }$\\
\bottomrule
\end{tabular}
\renewcommand{\arraystretch}{1}
\begin{tablenotes} 
\item[*] The numbers enclosed in parentheses correspond to the five rings presented in Table~\ref{tab:params}. 
\item {Note. -- } The listed values and errors correspond to the median value and the 68 per cent range of the posterior distributions, respectively. 
\end{tablenotes}
\end{threeparttable}
\end{table}

Another interesting feature is seen at the residual plot of Fig.~\ref{fig:stacked} close to the locations of peaks \#8, \#9 and \#10. First, the residual structure around \#8 indicates multiple peaks that are not resolved.  Second, large residuals are found before and after the peaks \#8 and \#10 respectively. These residuals are caused by the width of Lorentzian profiles used for describing peaks \#8 and \#10 that lead to excess emission over the data. Clearly the mathematical description could be improved by inserting two more Lorentzian lines. Higher resolution instruments like \textit{Chandra} could potentially identify more peaks in this range of angles that would correspond to layer distances between 0.4-0.7 kpc. We finally note that the outermost rings translate to layers at distances of only 74~pc. This is intriguing and highlights the power of X-ray tomography in providing distance measurements to dust layers even in regions of the Galaxy that cannot be mapped as accurately by other techniques. 

\begin{figure}
\includegraphics[width=0.49\textwidth, trim = 0 75 0 0]{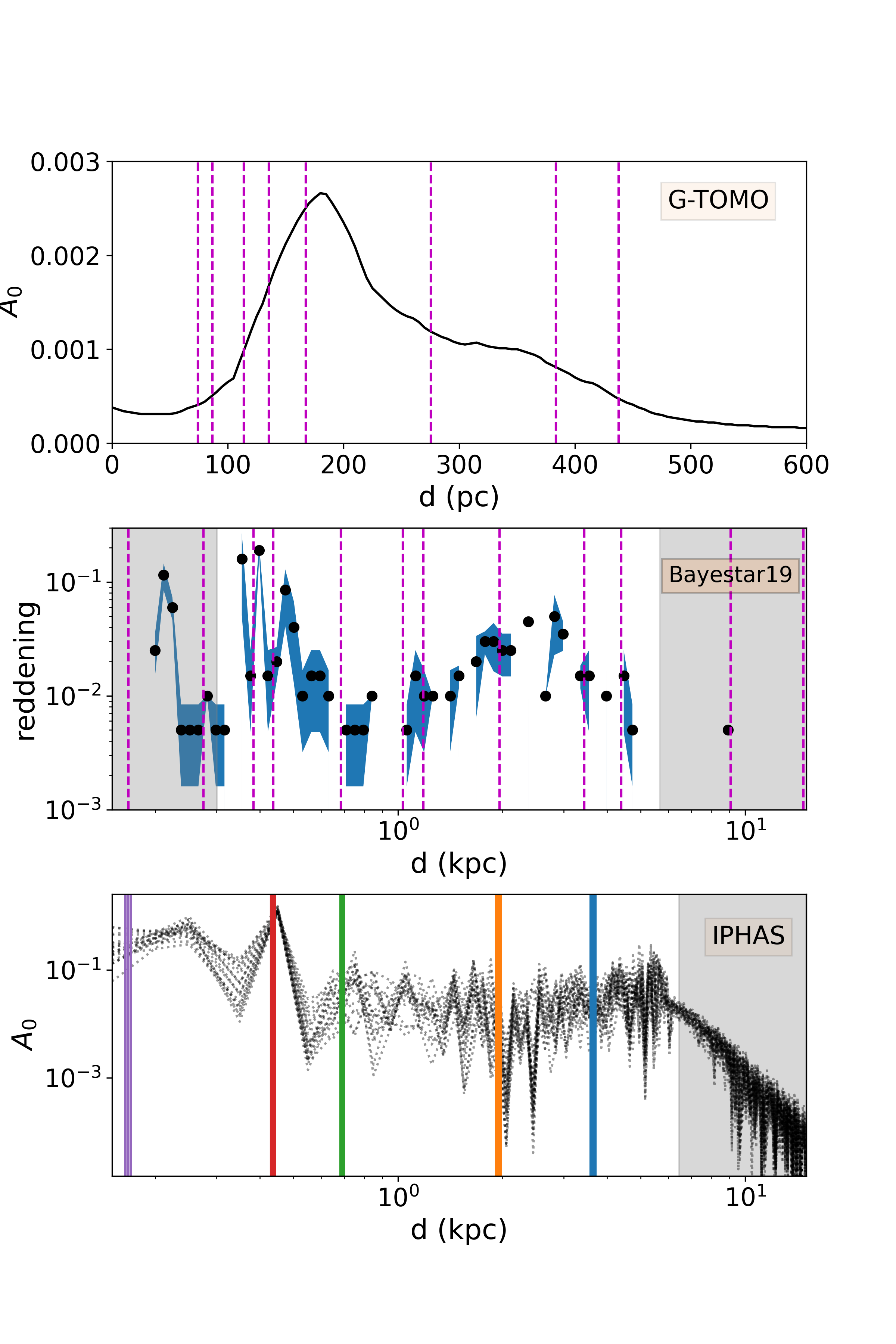}
\caption{Comparison of dust layer distances with various means of Galactic extinction measures in the direction of GRB~221007A.  \textit{Top panel:} Extinction profile (in terms of the monochromatic extinction at 5500 \AA) based on Gaia EDR3 and 2MASS data \citep{2022A&A...661A.147L}.  \textit{Middle panel:} Posterior distribution of reddening based on {\tt Bayestar19} 3D extinction maps \citep{2015ApJ...810...25G,2019ApJ...887...93G}. \textit{Bottom panel}: Posterior distribution of the mean extinction at 5495 \AA \, based on IPHAS photometry \citep{2014MNRAS.443.2907S}.  The reliability range of the extinction estimates is indicated with grey shaded areas. Vertical lines indicate the location of the dust layers found in the stacked data (top and middle panels) or in the individual data (bottom panel). } \label{fig:ebv}
\end{figure}

\section{Discussion}
\label{sec:discussion}
\subsection{Comparison with other probes of dust}
The dust content in our Galaxy is typically studied via reddening of starlight and CO emission from cold gas, while dust-scattering rings offer a new dimension to the above.
After estimating the location of the dust layers we can compare their position with the Galactic extinction profile along the line of sight due to dust attenuation. We first use the data from {\tt Bayestar19} 3D maps, i.e. the latest version of the Dust Map based on Gaia, Pan-STARRS 1, and 2MASS data \citep{2015ApJ...810...25G,2019ApJ...887...93G}. Given the probabilistic nature of the maps we extract 1000 random samples for the direction of our source and estimate the median and 68 per cent confidence range for the differential reddening value.  We note that the output values of the 3D map are given in arbitrary units; we refer the reader to \cite{2015ApJ...810...25G,2019ApJ...887...93G} for a description of the conversion to $E(B-V)$ or extinction $A$ in a specific pass band. We also extract the mean extinction (at the reference wavelength of 5495 \AA) along the direction of the burst from \cite{2014MNRAS.443.2907S} who derived the 3D map of extinction in the northern Galactic plane ($|b|<5^{\rm o}$) using IPHAS DR2 photometry.
The IPHAS map provides cumulative extinction values which for the direction of the system correspond to about 4 magnitudes (up to a distance of $\sim 6$~kpc where the results are trustworthy). The extinction can also be used as a proxy for hydrogen column density according to $N_{\rm H}=2.21\times10^{21}$~$A_{\rm V}$~cm$^{-2}$ assuming solar metallicity \citep{2009MNRAS.400.2050G}. The estimated column density is $N_{\rm H} \sim 0.9 \times 10^{22}$~cm$^{-2}$ (assuming $A_0\approx A_{\rm V}$). Both extinction maps discussed so far have low resolution to smaller distances (within 1 kpc). Therefore, to obtain a better picture of the local extinction profile we use the updated Gaia-2MASS 3D maps of Galactic interstellar dust \citep{2022A&A...661A.147L}, which are available via the {\tt G-TOMO} online tool in the EXPLORE website\footnote{\url{https://explore-platform.eu}}.  

The results are shown in Fig.~\ref{fig:ebv} where the vertical lines indicate the locations of the dust layers derived from the analysis of individual XRT datasets (bottom panel) and of the stacked image (top and middle panels). There is some agreement between the inferred distances for the nearby layers ($\lesssim$ 1~kpc) and the positions of larger $A_{\rm 0}$ (and thus $N_{\rm H}$) values. Estimates for the amount of dust from extinction measurements are limited to smaller distances, since the amount of stars and the accuracy of photometry decreases as we move to the outskirts of the Galaxy. For instance, the extinction estimates from IPHAS are not trustworthy beyond $\sim 6~$kpc (see shaded regions in panels of Fig.~\ref{fig:ebv}). Meanwhile, X-ray scattering by dust closer to us produces rings with larger angular sizes that are more difficult to detect due to e.g. lower intensity. Overall, performing an X-ray tomography of the Galaxy via dust scattering echoes favours the detection of layers at larger distances (the scattering angle is smaller and the ring intensity larger), thus complementing photometric techniques for dust mapping. 

\begin{figure}
\setbox1=\hbox{\includegraphics[width=0.48\textwidth]{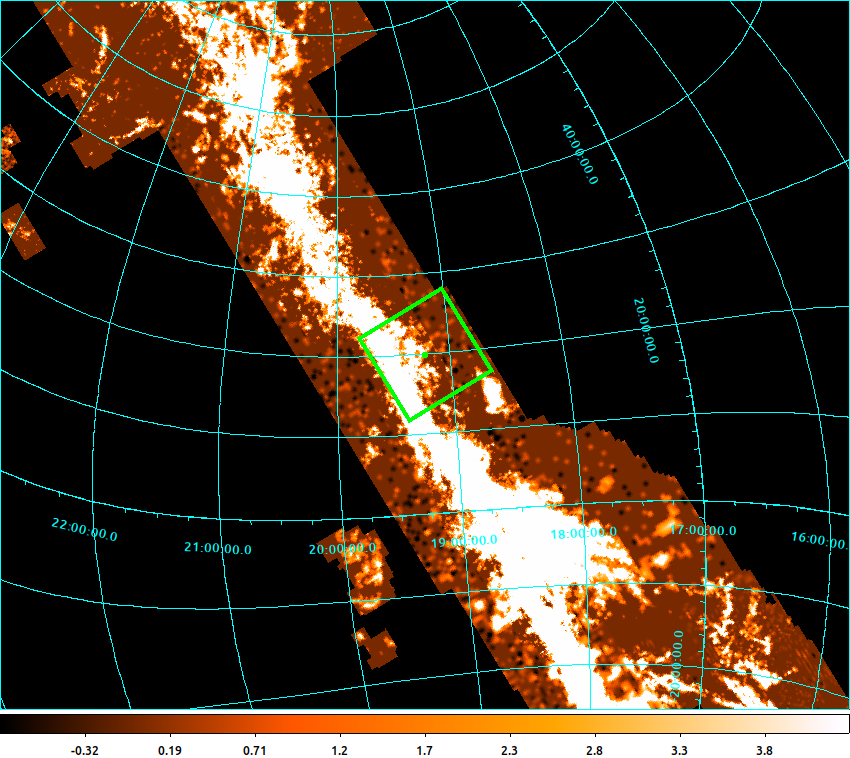}}
\includegraphics[width=0.48\textwidth]{fig7.png}\llap{\raisebox{4.3cm}{\includegraphics[height=3.3cm]{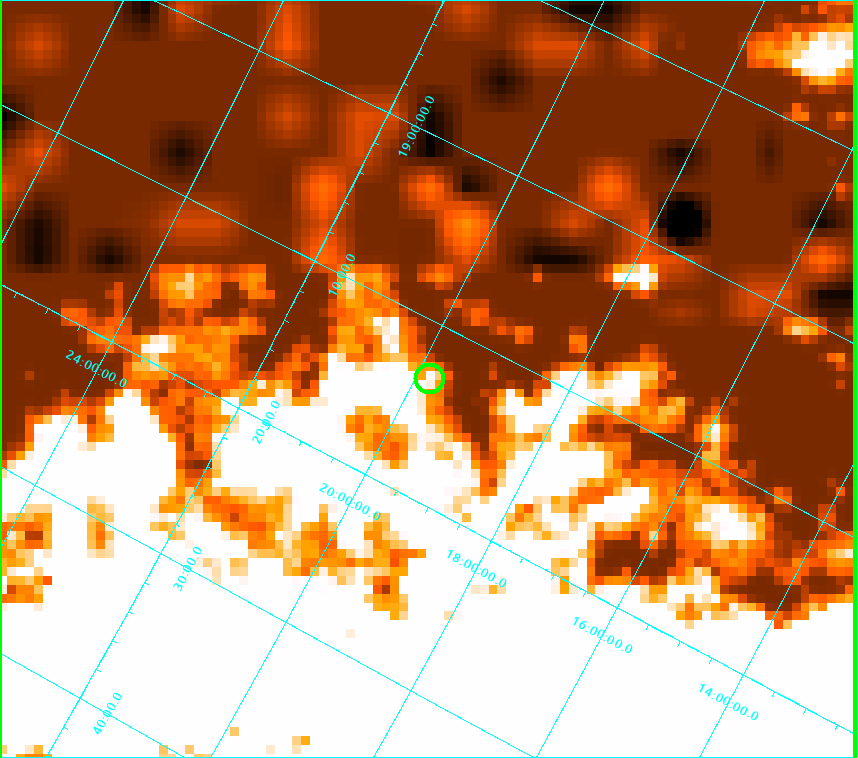}}}
\caption{
Velocity-integrated spatial CO sky-map with brighter colours corresponding to larger values \citep{2001ApJ...547..792D}. Inset plot shows a zoom-in version of the central image. A circle with angular size of 12\arcmin~ marks the location of GRB 221009A, its size is comparable to the radius of the observed rings, the size is also comparable to the CO map resolution (i.e. pixel size).
}
\label{fig:co}
\end{figure}

\begin{figure*}
\includegraphics[width=0.99\textwidth]{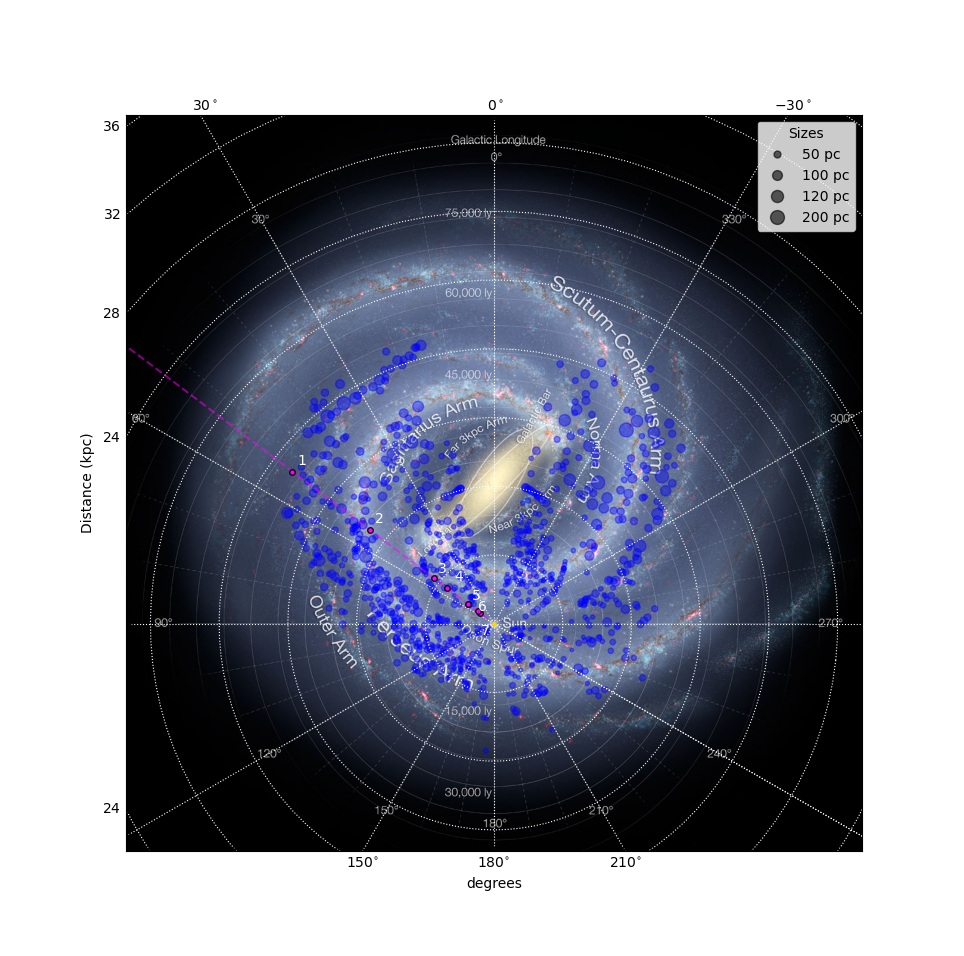}
\vspace{-1cm}
\caption{Location of massive molecular clouds (blue circles) in the Galactic plane as obtained from CO measurements \citep{2016ApJ...822...52R}. The size of the markers corresponds to the actual cloud size. The direction of GRB 221007A is marked with a magenta dashed line. The most prominent dust layers at distances of $\sim$1.03, 1.18, 1.96, 3.44, 4.40, 9.07 and 14.7 kpc are marked with magenta circles,  and Arabic numbers corresponding to the rings 7, 6, 5, 4, 3, 2, 1 respectively (see Table \ref{tab:params-stacked}). The Sun's location is marked with a yellow circle. Background illustration of the Milky Way reflecting the Galactic structure [Image credit: NASA/JPL-Caltech/ESO/R. Hurt] }
\label{fig:MW}
\end{figure*}

To better visualize the direction of the source compared to the Galactic plane we show in Fig.~\ref{fig:co} its location in the sky on top of the velocity-integrated spatial CO map \citep{2001ApJ...547..792D}. 
The map provides radial velocities that could be de-projected and translated to distances. However, this is far from an easy task, which does not always result in a unique solution for the distance of the CO emitting gas, but can yield instead a near and a far distance solution.  \citet{2016ApJ...822...52R} used a dendrogram-based decomposition of the \cite{2001ApJ...547..792D} survey and constructed a catalog of 1064 massive molecular clouds
throughout the Galactic plane. These massive cold clouds are another tracer of dust concentrations in our Galaxy. In Fig. \ref{fig:MW} we project the catalog of the molecular clouds  (blue points) onto an illustration of the Milky way and compare those with the dust layers as inferred from the rings at distances of $\sim$1.03, 1.18, 1.96, 3.44, 4.40, 9.07 and 14.7 kpc (magenta points).
We did not identify any dust layers between 5 and 9 kpc through the ring analysis, which agrees with the paucity in the molecular cloud distribution and the gap between the Sagittarius and Perseus spiral arms (Fig.~\ref{fig:MW}). We note that the molecular clouds are confined to the Galactic plane ($|b| \lesssim 2^{\rm o}$) with radii of the order of 100~pc,  
while our line of sight probes dust distributed above the plane ($b=4.32^{\rm o}$).  Even though a direct connection of the cloud and layer distributions cannot be made, it is plausible that the dust extending above the plane follows a similar distribution as the one probed by the clouds.  

\begin{figure}
\centering
\includegraphics[width=0.45\textwidth]{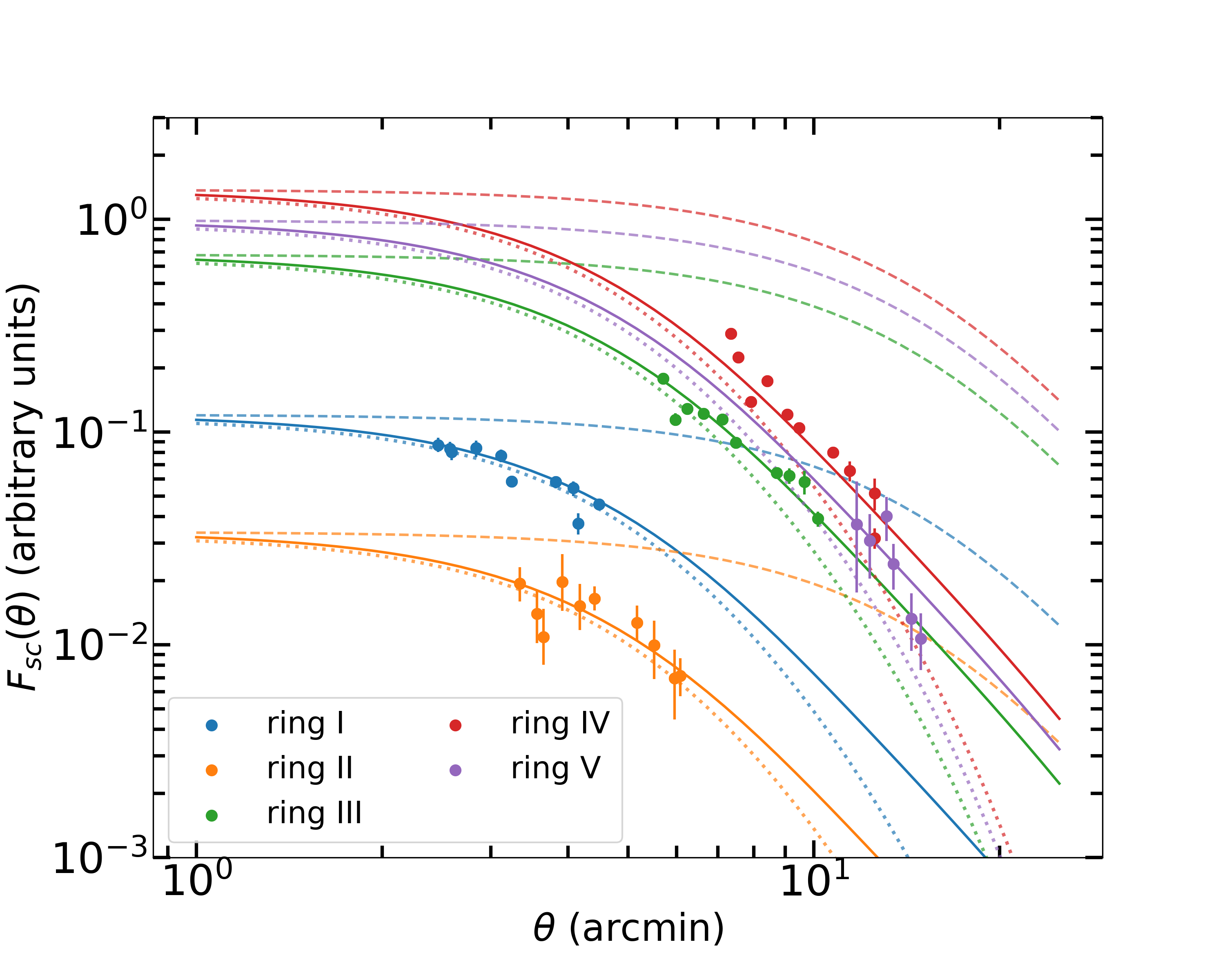}
\caption{Scattered flux of X-ray rings  (in arbitrary units) integrated over the angular extent and plotted as a function of the angular size $\theta$ (coloured symbols). Theoretical expectations based on the simplest grain model are overplotted for indicative parameter values: $a_{\max} = 0.3~{\mu m}, a_{\min}=a_{\max}/10$ (solid lines), $a_{\max} = 0.1~{\mu m}, a_{\min}=a_{\max}/10$ (dashed lines), and $a_{\max} = 0.3~{\mu m}, a_{\min}=a_{\max}/3$ (dotted lines). In all cases, $q=4$. Theoretical curves (for each parameter set) are normalized to the same value at 1~arcmin.
}
\label{fig:Fsc}
\end{figure}

\subsection{Scattered X-ray intensity}
The evolution of the X-ray scattered intensity with time (or angular size) is associated with the dust grain properties.
The X-ray flux of a ring with angular size $\theta$, which is produced by scattering of a infinitesimally short duration burst of X-rays with fluence $S_{\rm X}(E)$ by dust at distance $x_{\rm i}$, can be written as \citep[for details see][]{2016MNRAS.455.4426V} 
\begin{equation}
F_{\rm sc}(E,\theta) = \frac{C_{\rm i} N_{\rm d, i} S_{\rm X}(E) }{x_{\rm i}(1-x_{\rm i}) }\int_{\tilde{a}_{\min}}^{\tilde{a}_{\max}} {\rm d}\tilde{a} \, \tilde{a}^{6-q}\exp\left(-\frac{\theta^2}{2 (1-x_{\rm i})^2 \Theta^2(E,\tilde{a})}\right)
\label{eq:Fsc}
\end{equation}
where $C_{\rm i}$ is a normalization constant that depends on the metallicity and mass density of dust in layer $i$ and is of order unity for typical parameters \citep[e.g.][]{2016MNRAS.455.4426V} and $N_{\rm d, i}$ is the dust column density of the $i$-th layer. The integral of the differential scattering cross section, which is modelled using the Rayleigh-Gans approximation~\citep[e.g.][]{Mauche1986}, is performed over a power-law grain size distribution with slope $q$~\citep[][]{MRN1977}; here $\tilde{a}$ is the grain size in $\mu m$ and $\Theta$ is the typical angular size of a ring produced via scattering of 1~keV photons on grains with radius 0.1~$\mu m$, 
\begin{equation}
\Theta = 10.4 \, {\rm arcmin} \left(\frac{1~\rm keV}{E} \right) \left( \frac{0.1~\rm \mu m}{a}\right).
\label{eq:theta2}
\end{equation} 
For photon energies $E > 1$~keV, as those considered in this paper, Eq.~(\ref{eq:Fsc}) is valid for $\tilde{a} \ll 1$. Most photons in the analyzed XRT images have energies between 1 and 2 keV. We therefore integrate the flux given by Eq.~(\ref{eq:Fsc}) over this narrow band and perform a qualitative comparison to the scattered fluxes derived from the optimal angular-profile models of the rings (see Fig.~\ref{fig:radialP}). We model the prompt X-ray fluence as $S(E) \propto (E/E_{\rm pk})^{-\Gamma+1}$, where $E_{\rm pk}=1060$~keV is the observed peak energy of the prompt spectrum as estimated from KONUS-WIND \citep{GRB221009A_KONUS_GCN32668} and $\Gamma = 3/2$ is the photon index of the prompt GRB spectrum, assuming a fast-cooling synchrotron spectrum extending down to 1~keV \citep{2022arXiv221200766R}. 

The theoretical expectations for indicative dust parameters 
are shown in Fig.~\ref{fig:Fsc}. In all cases, we assume a power-law size distribution of grains with slope $q=4$ extending from $a_{\min}$ to $a_{\max}$. Solid lines correspond to $a_{\max}=0.3 \mu m, a_{\min}=a_{\max}/10$, dashed lines to $a_{\max}=0.1 \mu m, a_{\min}=a_{\max}/10$, and dotted lines to $a_{\max}=0.3 \mu m, a_{\min}=a_{\max}/3$. We do not determine the normalization parameter for each dust layer, $\tilde{C}_{\rm i}=C_{\rm i}N_{\rm d,i}$, as we are interested in the relative ratio of the fluxes. Even without fitting the model to the data we can draw some useful conclusions. First, the maximum grain size cannot be much smaller than $0.3~\mu m$. For example, $F_{\rm sc}(\theta)$ would be almost constant for $\theta \lesssim 10$~arcmin if $a_{\max}=0.1~\mu m$ in contradiction to the data (see dashed lines). The smooth turnover of $F_{\rm sc}(\theta)$ is related to the exponential cutoff in the scattering cross section (see Eq.~(\ref{eq:Fsc})), and occurs approximately at $\Theta(\bar{E}, a_{\max})$, which is $\simeq 2.5$~arcmin for a mean photon energy $\bar{E}=1.5$~keV and $a_{\max}=0.3~\mu m$ -- see Eq.~(\ref{eq:theta2}). Second, the minimum grain size cannot be easily constrained because of the small dynamic range of the ring angular sizes. In general, the scattered X-ray flux follows a power law in angle, with a slope depending on $q$, and an extent determined roughly by $\Theta(\bar{E}, a_{\max})$ and $\Theta(\bar{E}, a_{\min})$ -- see e.g. green and red solid lines. As $a_{\min}$ approaches $a_{\max}$, however, the power-law segment of $F_{\rm sc}(\theta)$ becomes shorter, till the point that we start seeing the exponential cutoff of the scattering cross section for grains of typical size $a_{\min}\sim a_{\max}$ (compare solid and dotted lines). Grain distributions with $a_{\min} \ll a_{\max}$ or $a_{\min} \sim a_{\max}$ are compatible with the data for rings I, II, and V. In fact, the scattered flux of the fifth ring would be better described by a model of grains with similar size instead of an extended power-law distribution (compare purple solid and dotted lines). 
Third, grain distributions with $q\sim 3.5-4$ and $a_{\min} \ll a_{\max}$ can produce the observed power-law decline of the scattered flux with angular size for rings III and V. Lastly, the relative normalizations for the dust layers are $\tilde{C}_{\rm I}:\tilde{C}_{\rm II}:\tilde{C}_{\rm III}:\tilde{C}_{\rm IV}:\tilde{C}_{\rm V} = 1:0.15:0.9:1.4:0.3$. The relative normalizations can be used to order the dust scattering production sites in terms of increasing optical depth or amount of dust contained in each layer, with the fourth layer (at 0.44~kpc) being the one with the largest dust content.

Prompt X-ray scattering by dust in the GRB host galaxy can also be imprinted in the X-ray afterglow emission \citep[e.g.][]{1998ApJ...507..300K, 2007ApJ...660.1319S}. For instance, the strong hard-to-soft evolution of the X-ray emission observed in the afterglow of the ultra-long GRB 130925A could be explained by this phenomenon \cite{2014MNRAS.444..250E}. The X-ray echoes of GRB 221009A are instead produced via scattering of prompt X-ray photons by dust in our Galaxy, as demonstrated in Sec.~\ref{sec:results}. Still, spectral softening with time is also expected. However, the X-ray afterglow of GRB 221009A shows no evidence for strong spectral evolution with a photon index close to -2 for about two decades in time\footnote{ \url{https://www.swift.ac.uk/burst_analyser/01126853/}}. In the small-angle scattering approximation, the scattered flux shows a shallow decline with time, i.e. $t^{-1/4}$ -- see e.g.~ Eq.~(3) in \cite{2007ApJ...660.1319S}. A steeper decline approaching $t^{-2}$ is expected after $t \gtrsim 1.6\times 10^5~{\rm s} \left(E/{1~{\rm keV}}\right)^{-2}\left(a/0.1~\mu m\right)^{-2}\left(d_\ell/100~{\rm pc}\right)(1+z_{\rm s})$. Therefore, a transition from a shallow decay to a steeper decline in the X-ray scattered flux would be expected somewhere between $6.5\times10^4$~s and $\sim 1.5\times10^6$~s  for layers at distances between 0.4~kpc and 9.6~kpc, respectively.  The XRT light curve shows no evidence of such transition, and its flux decays almost as a single power law (with slope $\sim -1.6$ for $t\gtrsim 10^4$~s after the GBM trigger. Comparison of dust-scattering models to the XRT afterglow light curve might help to constrain the dust column density of each layer and estimate the contribution of the scattered flux to the intrinsic non-thermal emission from the GRB blast wave.

\section{Conclusions}\label{sec:summary}

In this paper we have analyzed publicly available \swift-XRT data that were obtained within a few days after the detection of GRB~221009A. We constructed angular profiles of photons with energies above 1 keV from individual XRT images, and identified the most prominent peaks. By modelling their temporal evolution over a course of several days we were able to determine the time of the X-ray burst and the distances of five intervening dust layers. Complementary analysis of the stacked XRT image (scaled to a reference time of two days after the burst) revealed a richer angular structure with 16 peaks due to the increased photon statistics.  The main conclusions of our work are the following:
\begin{itemize}
    \item The expansion of the five more prominent peaks in the time-resolved angular profiles yields the time of the X-ray burst, which is consistent with the GBM trigger (i.e. the prompt X-ray spectrum should extend to 1~keV). 
    \item Analysis of the stacked image reveals extra features and increases the number of potential dust concentrations along the line of sight to at least 16, spanning from 0.07~kpc to 15 kpc. This is this the largest distance range probed by X-ray scattering echoes so far.
    \item Locations of dust layers are generally consistent with local maxima of the radial extinction profile, while the absence of dust layers between 5 and 9 kpc coincides with the gap between the Sagittarius and Perseus spiral arms. 
    \item The evolution of the scattered X-ray flux (for the five more prominent rings) with angular size is consistent with scattering by dust grains having a power-law size distribution with slope $q\sim 3.5-4$ and maximum grain size of $0.3 \mu m$. For the closest layer to us, the minimum grain size could be comparable to $a_{\max}$. 
\end{itemize}

\section*{Acknowledgements}
We thank the referee for comments that helped to improve the manuscript. We are also grateful to Dr. Andrea Tiengo for identifying a typo in Eq.~(3) and for useful discussions. The authors acknowledge support by H.F.R.I. through the project ASTRAPE (Project ID 7802) and the project UNTRAPHOB (Project ID 3013). M.P. also acknowledges support from the MERAC Fondation through the project THRILL.

\section*{Data availability}
X-ray data are available through the High Energy Astrophysics Science Archive Research Center: \url{heasarc.gsfc.nasa.gov}. 
The python notebooks used for the X-ray image analysis and the radial profile fitting will be made available upon reasonable request to the authors.


\bibliographystyle{mnras}
\bibliography{bibliography}

\begin{thebibliography}{}
\makeatletter
\relax
\def\mn@urlcharsother{\let\do\@makeother \do\$\do\&\do\#\do\^\do\_\do\%\do\~}
\def\mn@doi{\begingroup\mn@urlcharsother \@ifnextchar [ {\mn@doi@}
  {\mn@doi@[]}}
\def\mn@doi@[#1]#2{\def\@tempa{#1}\ifx\@tempa\@empty \href
  {http://dx.doi.org/#2} {doi:#2}\else \href {http://dx.doi.org/#2} {#1}\fi
  \endgroup}
\def\mn@eprint#1#2{\mn@eprint@#1:#2::\@nil}
\def\mn@eprint@arXiv#1{\href {http://arxiv.org/abs/#1} {{\tt arXiv:#1}}}
\def\mn@eprint@dblp#1{\href {http://dblp.uni-trier.de/rec/bibtex/#1.xml}
  {dblp:#1}}
\def\mn@eprint@#1:#2:#3:#4\@nil{\def\@tempa {#1}\def\@tempb {#2}\def\@tempc
  {#3}\ifx \@tempc \@empty \let \@tempc \@tempb \let \@tempb \@tempa \fi \ifx
  \@tempb \@empty \def\@tempb {arXiv}\fi \@ifundefined
  {mn@eprint@\@tempb}{\@tempb:\@tempc}{\expandafter \expandafter \csname
  mn@eprint@\@tempb\endcsname \expandafter{\@tempc}}}

\bibitem[\protect\citeauthoryear{{A. Ursi et al.}}{{A. Ursi et
  al.}}{2022}]{GRB221009A_AGILE_GCN32650}
{A. Ursi et al.} 2022, GCN Circ. 32650

\bibitem[\protect\citeauthoryear{{Ajello} et~al.,}{{Ajello}
  et~al.}{2019}]{fermilat2nd}
{Ajello} M.,  et~al., 2019, \mn@doi [\apj] {10.3847/1538-4357/ab1d4e}, \href
  {https://ui.adsabs.harvard.edu/abs/2019ApJ...878...52A} {878, 52}

\bibitem[\protect\citeauthoryear{{Beardmore}, {Willingale}, {Kuulkers},
  {Altamirano}, {Motta}, {Osborne}, {Page}  \& {Sivakoff}}{{Beardmore}
  et~al.}{2016}]{2016MNRAS.462.1847B}
{Beardmore} A.~P.,  {Willingale} R.,  {Kuulkers} E.,  {Altamirano} D.,  {Motta}
  S.~E.,  {Osborne} J.~P.,  {Page} K.~L.,   {Sivakoff} G.~R.,  2016, \mn@doi
  [\mnras] {10.1093/mnras/stw1753}, \href
  {https://ui.adsabs.harvard.edu/abs/2016MNRAS.462.1847B} {462, 1847}

\bibitem[\protect\citeauthoryear{{Burrows} et~al.,}{{Burrows}
  et~al.}{2005}]{2005SSRv..120..165B}
{Burrows} D.~N.,  et~al., 2005, \mn@doi [\ssr] {10.1007/s11214-005-5097-2},
  \href {https://ui.adsabs.harvard.edu/abs/2005SSRv..120..165B} {120, 165}

\bibitem[\protect\citeauthoryear{Chiang \& Dermer}{Chiang \&
  Dermer}{1999}]{CD1999}
Chiang J.,  Dermer C.~D.,  1999, \mn@doi [The Astrophysical Journal]
  {10.1086/306789}, 512, 699

\bibitem[\protect\citeauthoryear{{Costantini} \& {Corrales}}{{Costantini} \&
  {Corrales}}{2022}]{2022arXiv220905261C}
{Costantini} E.,  {Corrales} L.,  2022, arXiv e-prints, \href
  {https://ui.adsabs.harvard.edu/abs/2022arXiv220905261C} {p. arXiv:2209.05261}

\bibitem[\protect\citeauthoryear{{D. Frederiks et al.}}{{D. Frederiks et
  al.}}{2022}]{GRB221009A_KONUS_GCN32668}
{D. Frederiks et al.} 2022, GCN Circ. 32668

\bibitem[\protect\citeauthoryear{{Dame}, {Hartmann}  \& {Thaddeus}}{{Dame}
  et~al.}{2001}]{2001ApJ...547..792D}
{Dame} T.~M.,  {Hartmann} D.,   {Thaddeus} P.,  2001, \mn@doi [\apj]
  {10.1086/318388}, \href
  {https://ui.adsabs.harvard.edu/abs/2001ApJ...547..792D} {547, 792}

\bibitem[\protect\citeauthoryear{{Dichiara}, {Gropp}, {Kennea}, {Kuin}, {Lien},
  {Marshall}, {Tohuvavohu}  \& {Williams}}{{Dichiara}
  et~al.}{2022}]{2022ATel15650....1D}
{Dichiara} S.,  {Gropp} J.~D.,  {Kennea} J.~A.,  {Kuin} N.~P.~M.,  {Lien}
  A.~Y.,  {Marshall} F.~E.,  {Tohuvavohu} A.,   {Williams} M.~A.,  2022, The
  Astronomer's Telegram, \href
  {https://ui.adsabs.harvard.edu/abs/2022ATel15650....1D} {15650, 1}

\bibitem[\protect\citeauthoryear{{Evans} et~al.,}{{Evans}
  et~al.}{2007}]{2007A&A...469..379E}
{Evans} P.~A.,  et~al., 2007, \mn@doi [\aap] {10.1051/0004-6361:20077530},
  \href {https://ui.adsabs.harvard.edu/abs/2007A&A...469..379E} {469, 379}

\bibitem[\protect\citeauthoryear{{Evans} et~al.,}{{Evans}
  et~al.}{2009}]{2009MNRAS.397.1177E}
{Evans} P.~A.,  et~al., 2009, \mn@doi [\mnras]
  {10.1111/j.1365-2966.2009.14913.x}, \href
  {https://ui.adsabs.harvard.edu/abs/2009MNRAS.397.1177E} {397, 1177}

\bibitem[\protect\citeauthoryear{{Evans} et~al.,}{{Evans}
  et~al.}{2014}]{2014MNRAS.444..250E}
{Evans} P.~A.,  et~al., 2014, \mn@doi [\mnras] {10.1093/mnras/stu1459}, \href
  {https://ui.adsabs.harvard.edu/abs/2014MNRAS.444..250E} {444, 250}

\bibitem[\protect\citeauthoryear{{Foreman-Mackey}, {Hogg}, {Lang}  \&
  {Goodman}}{{Foreman-Mackey} et~al.}{2013}]{2013PASP..125..306F}
{Foreman-Mackey} D.,  {Hogg} D.~W.,  {Lang} D.,   {Goodman} J.,  2013, \mn@doi
  [\pasp] {10.1086/670067}, \href
  {https://ui.adsabs.harvard.edu/abs/2013PASP..125..306F} {125, 306}

\bibitem[\protect\citeauthoryear{{Green} et~al.,}{{Green}
  et~al.}{2015}]{2015ApJ...810...25G}
{Green} G.~M.,  et~al., 2015, \mn@doi [\apj] {10.1088/0004-637X/810/1/25},
  \href {https://ui.adsabs.harvard.edu/abs/2015ApJ...810...25G} {810, 25}

\bibitem[\protect\citeauthoryear{{Green}, {Schlafly}, {Zucker}, {Speagle}  \&
  {Finkbeiner}}{{Green} et~al.}{2019}]{2019ApJ...887...93G}
{Green} G.~M.,  {Schlafly} E.,  {Zucker} C.,  {Speagle} J.~S.,   {Finkbeiner}
  D.,  2019, \mn@doi [\apj] {10.3847/1538-4357/ab5362}, \href
  {https://ui.adsabs.harvard.edu/abs/2019ApJ...887...93G} {887, 93}

\bibitem[\protect\citeauthoryear{{G{\"u}ver} \& {{\"O}zel}}{{G{\"u}ver} \&
  {{\"O}zel}}{2009}]{2009MNRAS.400.2050G}
{G{\"u}ver} T.,  {{\"O}zel} F.,  2009, \mn@doi [\mnras]
  {10.1111/j.1365-2966.2009.15598.x}, \href
  {https://ui.adsabs.harvard.edu/abs/2009MNRAS.400.2050G} {400, 2050}

\bibitem[\protect\citeauthoryear{{Heinz} et~al.,}{{Heinz}
  et~al.}{2015}]{2015ApJ...806..265H}
{Heinz} S.,  et~al., 2015, \mn@doi [\apj] {10.1088/0004-637X/806/2/265}, \href
  {https://ui.adsabs.harvard.edu/abs/2015ApJ...806..265H} {806, 265}

\bibitem[\protect\citeauthoryear{{Heinz}, {Corrales}, {Smith}, {Brandt},
  {Jonker}, {Plotkin}  \& {Neilsen}}{{Heinz}
  et~al.}{2016}]{2016ApJ...825...15H}
{Heinz} S.,  {Corrales} L.,  {Smith} R.,  {Brandt} W.~N.,  {Jonker} P.~G.,
  {Plotkin} R.~M.,   {Neilsen} J.,  2016, \mn@doi [\apj]
  {10.3847/0004-637X/825/1/15}, \href
  {https://ui.adsabs.harvard.edu/abs/2016ApJ...825...15H} {825, 15}

\bibitem[\protect\citeauthoryear{{Hinshaw} et~al.,}{{Hinshaw}
  et~al.}{2013}]{wmap9}
{Hinshaw} G.,  et~al., 2013, \mn@doi [\apjs] {10.1088/0067-0049/208/2/19},
  \href {https://ui.adsabs.harvard.edu/abs/2013ApJS..208...19H} {208, 19}

\bibitem[\protect\citeauthoryear{{Jin}, {Ponti}, {Haberl}  \& {Smith}}{{Jin}
  et~al.}{2017}]{2017MNRAS.468.2532J}
{Jin} C.,  {Ponti} G.,  {Haberl} F.,   {Smith} R.,  2017, \mn@doi [\mnras]
  {10.1093/mnras/stx653}, \href
  {https://ui.adsabs.harvard.edu/abs/2017MNRAS.468.2532J} {468, 2532}

\bibitem[\protect\citeauthoryear{{Jin}, {Ponti}, {Haberl}, {Smith}  \&
  {Valencic}}{{Jin} et~al.}{2018}]{2018MNRAS.477.3480J}
{Jin} C.,  {Ponti} G.,  {Haberl} F.,  {Smith} R.,   {Valencic} L.,  2018,
  \mn@doi [\mnras] {10.1093/mnras/sty869}, \href
  {https://ui.adsabs.harvard.edu/abs/2018MNRAS.477.3480J} {477, 3480}

\bibitem[\protect\citeauthoryear{{Jin}, {Ponti}, {Li}  \& {Bogensberger}}{{Jin}
  et~al.}{2019}]{2019ApJ...875..157J}
{Jin} C.,  {Ponti} G.,  {Li} G.,   {Bogensberger} D.,  2019, \mn@doi [\apj]
  {10.3847/1538-4357/ab11d1}, \href
  {https://ui.adsabs.harvard.edu/abs/2019ApJ...875..157J} {875, 157}

\bibitem[\protect\citeauthoryear{{Karaferias}, {Vasilopoulos}, {Petropoulou},
  {Jenke}, {Wilson-Hodge}  \& {Malacaria}}{{Karaferias}
  et~al.}{2022}]{2022MNRAS.tmp.3021K}
{Karaferias} A.~S.,  {Vasilopoulos} G.,  {Petropoulou} M.,  {Jenke} P.~A.,
  {Wilson-Hodge} C.~A.,   {Malacaria} C.,  2022, \mn@doi [\mnras]
  {10.1093/mnras/stac3208}, \href
  {https://ui.adsabs.harvard.edu/abs/2022MNRAS.tmp.3021K} {}

\bibitem[\protect\citeauthoryear{{Klose}}{{Klose}}{1994}]{Klose1994}
{Klose} S.,  1994, \mn@doi [\apjl] {10.1086/187226}, \href
  {http://adsabs.harvard.edu/abs/1994ApJ...423L..23K} {423, L23}

\bibitem[\protect\citeauthoryear{{Klose}}{{Klose}}{1998}]{1998ApJ...507..300K}
{Klose} S.,  1998, \mn@doi [\apj] {10.1086/306335}, \href
  {https://ui.adsabs.harvard.edu/abs/1998ApJ...507..300K} {507, 300}

\bibitem[\protect\citeauthoryear{{Kumar} \& {Zhang}}{{Kumar} \&
  {Zhang}}{2015}]{2015PhR...561....1K}
{Kumar} P.,  {Zhang} B.,  2015, \mn@doi [\physrep]
  {10.1016/j.physrep.2014.09.008}, \href
  {https://ui.adsabs.harvard.edu/abs/2015PhR...561....1K} {561, 1}

\bibitem[\protect\citeauthoryear{{Lallement}, {Vergely}, {Babusiaux}  \&
  {Cox}}{{Lallement} et~al.}{2022}]{2022A&A...661A.147L}
{Lallement} R.,  {Vergely} J.~L.,  {Babusiaux} C.,   {Cox} N.~L.~J.,  2022,
  \mn@doi [\aap] {10.1051/0004-6361/202142846}, \href
  {https://ui.adsabs.harvard.edu/abs/2022A&A...661A.147L} {661, A147}

\bibitem[\protect\citeauthoryear{{Lamer}, {Schwope}, {Predehl}, {Traulsen},
  {Wilms}  \& {Freyberg}}{{Lamer} et~al.}{2021}]{2021A&A...647A...7L}
{Lamer} G.,  {Schwope} A.~D.,  {Predehl} P.,  {Traulsen} I.,  {Wilms} J.,
  {Freyberg} M.,  2021, \mn@doi [\aap] {10.1051/0004-6361/202039757}, \href
  {https://ui.adsabs.harvard.edu/abs/2021A&A...647A...7L} {647, A7}

\bibitem[\protect\citeauthoryear{{Mathis}, {Rumpl}  \& {Nordsieck}}{{Mathis}
  et~al.}{1977}]{MRN1977}
{Mathis} J.~S.,  {Rumpl} W.,   {Nordsieck} K.~H.,  1977, \mn@doi [\apj]
  {10.1086/155591}, \href {http://adsabs.harvard.edu/abs/1977ApJ...217..425M}
  {217, 425}

\bibitem[\protect\citeauthoryear{{Mauche} \& {Gorenstein}}{{Mauche} \&
  {Gorenstein}}{1986}]{Mauche1986}
{Mauche} C.~W.,  {Gorenstein} P.,  1986, \mn@doi [\apj] {10.1086/163996}, \href
  {http://adsabs.harvard.edu/abs/1986ApJ...302..371M} {302, 371}

\bibitem[\protect\citeauthoryear{{Miceli} \& {Nava}}{{Miceli} \&
  {Nava}}{2022}]{2022Galax..10...66M}
{Miceli} D.,  {Nava} L.,  2022, \mn@doi [Galaxies] {10.3390/galaxies10030066},
  \href {https://ui.adsabs.harvard.edu/abs/2022Galax..10...66M} {10, 66}

\bibitem[\protect\citeauthoryear{{R. Pillera et al.}}{{R. Pillera et
  al.}}{2022}]{GRB221009A_LAT_GCN32658}
{R. Pillera et al.} 2022, GCN Circ. 32658

\bibitem[\protect\citeauthoryear{Rees \& Mészáros}{Rees \&
  Mészáros}{1992}]{RM1992}
Rees M.~J.,  Mészáros P.,  1992, \mn@doi [Monthly Notices of the Royal
  Astronomical Society] {10.1093/mnras/258.1.41P}, 258, 41P

\bibitem[\protect\citeauthoryear{{Refsdal}}{{Refsdal}}{1966}]{1966MNRAS.132..101R}
{Refsdal} S.,  1966, \mn@doi [\mnras] {10.1093/mnras/132.1.101}, \href
  {https://ui.adsabs.harvard.edu/abs/1966MNRAS.132..101R} {132, 101}

\bibitem[\protect\citeauthoryear{{Rice}, {Goodman}, {Bergin}, {Beaumont}  \&
  {Dame}}{{Rice} et~al.}{2016}]{2016ApJ...822...52R}
{Rice} T.~S.,  {Goodman} A.~A.,  {Bergin} E.~A.,  {Beaumont} C.,   {Dame}
  T.~M.,  2016, \mn@doi [\apj] {10.3847/0004-637X/822/1/52}, \href
  {https://ui.adsabs.harvard.edu/abs/2016ApJ...822...52R} {822, 52}

\bibitem[\protect\citeauthoryear{{Rudolph}, {Petropoulou}, {Winter}  \&
  {Bo{\v{s}}njak}}{{Rudolph} et~al.}{2022}]{2022arXiv221200766R}
{Rudolph} A.,  {Petropoulou} M.,  {Winter} W.,   {Bo{\v{s}}njak} {\v{Z}}.,
  2022, arXiv e-prints, \href
  {https://ui.adsabs.harvard.edu/abs/2022arXiv221200766R} {p. arXiv:2212.00766}

\bibitem[\protect\citeauthoryear{{S. Lesage et al.}}{{S. Lesage et
  al.}}{2022}]{GRB221009A_GBM_GCN32642}
{S. Lesage et al.} 2022, GCN Circ. 32642

\bibitem[\protect\citeauthoryear{{Sale} et~al.,}{{Sale}
  et~al.}{2014}]{2014MNRAS.443.2907S}
{Sale} S.~E.,  et~al., 2014, \mn@doi [\mnras] {10.1093/mnras/stu1090}, \href
  {https://ui.adsabs.harvard.edu/abs/2014MNRAS.443.2907S} {443, 2907}

\bibitem[\protect\citeauthoryear{{Savitzky} \& {Golay}}{{Savitzky} \&
  {Golay}}{1964}]{1964AnaCh..36.1627S}
{Savitzky} A.,  {Golay} M.~J.~E.,  1964, Analytical Chemistry, \href
  {https://ui.adsabs.harvard.edu/abs/1964AnaCh..36.1627S} {36, 1627}

\bibitem[\protect\citeauthoryear{{Shao} \& {Dai}}{{Shao} \&
  {Dai}}{2007}]{2007ApJ...660.1319S}
{Shao} L.,  {Dai} Z.~G.,  2007, \mn@doi [\apj] {10.1086/513139}, \href
  {https://ui.adsabs.harvard.edu/abs/2007ApJ...660.1319S} {660, 1319}

\bibitem[\protect\citeauthoryear{{Tiengo}, {Pintore}, {Mereghetti}  \&
  {Salvaterra}}{{Tiengo} et~al.}{2022}]{2022ATel15661....1T}
{Tiengo} A.,  {Pintore} F.,  {Mereghetti} S.,   {Salvaterra} R.,  2022, The
  Astronomer's Telegram, \href
  {https://ui.adsabs.harvard.edu/abs/2022ATel15661....1T} {15661, 1}

\bibitem[\protect\citeauthoryear{{Vasilopoulos} \&
  {Petropoulou}}{{Vasilopoulos} \& {Petropoulou}}{2016}]{2016MNRAS.455.4426V}
{Vasilopoulos} G.,  {Petropoulou} M.,  2016, \mn@doi [\mnras]
  {10.1093/mnras/stv2605}, \href
  {https://ui.adsabs.harvard.edu/abs/2016MNRAS.455.4426V} {455, 4426}

\bibitem[\protect\citeauthoryear{{Vaughan} et~al.,}{{Vaughan}
  et~al.}{2004}]{Vaughan2004}
{Vaughan} S.,  et~al., 2004, \mn@doi [\apjl] {10.1086/382785}, \href
  {http://adsabs.harvard.edu/abs/2004ApJ...603L...5V} {603, L5}

\bibitem[\protect\citeauthoryear{{Vianello}, {Tiengo}  \&
  {Mereghetti}}{{Vianello} et~al.}{2007}]{Vianello2007}
{Vianello} G.,  {Tiengo} A.,   {Mereghetti} S.,  2007, \mn@doi [\aap]
  {10.1051/0004-6361:20077968}, \href
  {http://adsabs.harvard.edu/abs/2007A%26A...473..423V} {473, 423}

\bibitem[\protect\citeauthoryear{{de Ugarte Postigo} et~al.,}{{de Ugarte
  Postigo} et~al.}{2022}]{2022GCN.32648....1D}
{de Ugarte Postigo} A.,  et~al., 2022, GRB Coordinates Network, \href
  {https://ui.adsabs.harvard.edu/abs/2022GCN.32648....1D} {32648, 1}

\makeatother
\end{thebibliography}


\begin{appendix}
\section{MCMC modelling of ring expansion}

\begin{figure*}
\centering
\includegraphics[width=0.9\textwidth]{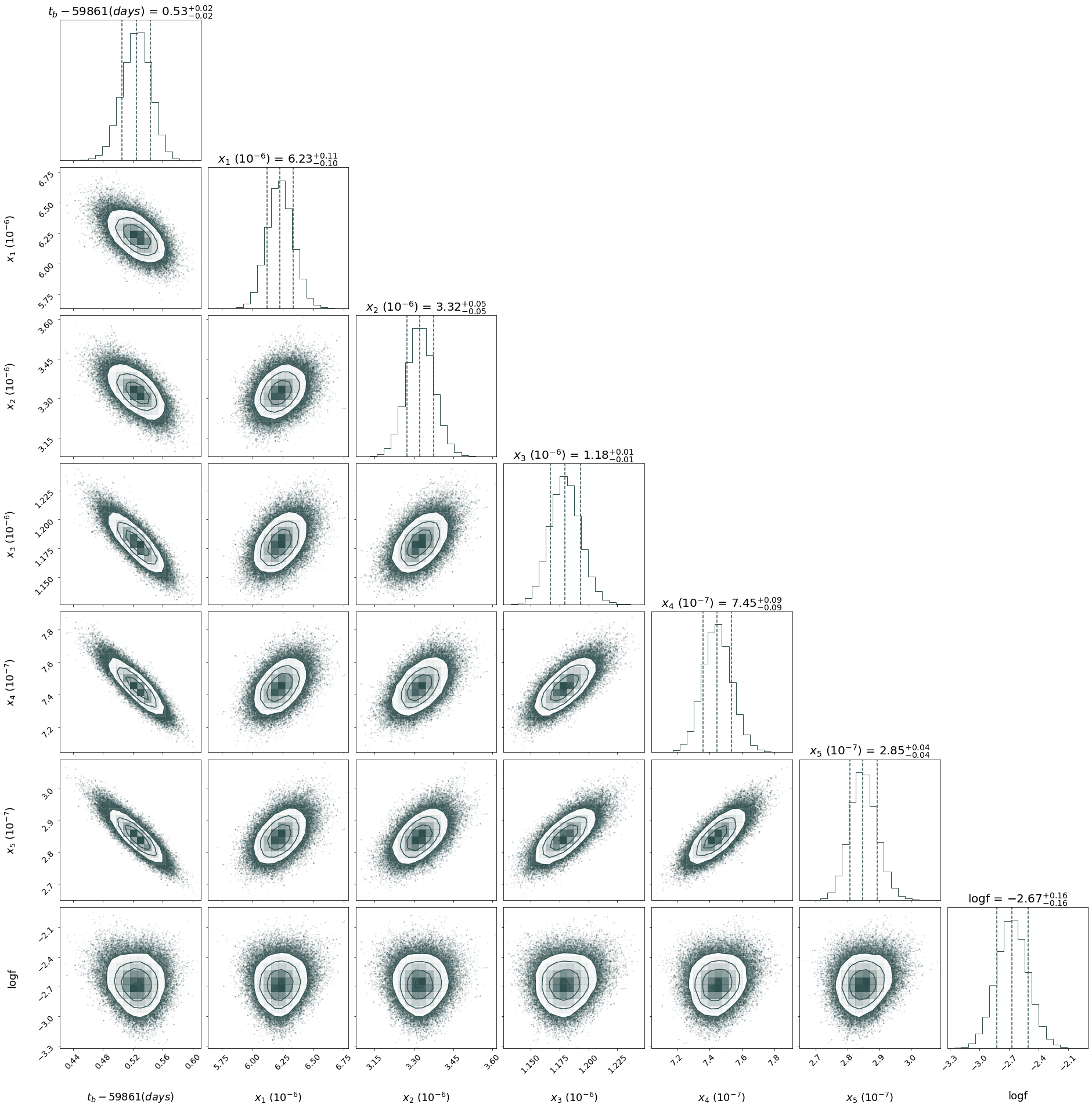} 
\caption{Corner plot showing the posterior distributions for the time of the burst $t_{\rm b}$ (in days since MJD~59861) and the distance of each dust layer normalized to the source distance, $x_i$.}
\label{fig:corner-2}
\end{figure*}

\renewcommand*{\arraystretch}{1.5} 
\onecolumn
\begin{table*}
\caption{Median values and 68 per cent confidence intervals derived from the posterior parameter distributions of the five prominent peaks identified in individual XRT datasets.}
\label{tab:optimal_params}
\centering

\begin{adjustbox}{width=1\textwidth} 
\begin{threeparttable}
\begin{tabular}{*{11}{c}}
\hline 
Lorentzian & \multicolumn{10}{c}{MJD}\\
Parameters & 59862.66 &
 59862.79 &
 59862.87 &
 59863.05 &
 59863.26 &
 59863.46 &
 59864.13 &
 59864.44 &
 59864.78 &
 59865.1  \\ 
\hline 
$a_{\rm L,1}$ &0.0352$_{-0.0026}^{+0.0028}$ & 0.0311$_{-0.0027}^{+0.0028}$ &0.0324$_{-0.0024}^{+0.0024}$  & 0.0296$_{-0.0027}^{+0.0025}$ & 0.0248$_{-0.0017}^{+0.0017}$ &0.0181$_{-0.0009}^{+0.0010}$ &  0.0152$_{-0.0010}^{+0.0010}$ &   0.0133$_{-0.0011}^{+0.0010}$ &  0.0090$_{-0.0010}^{+0.0010}$ &    0.0102$_{-0.0005}^{+0.0005}$  \\
$b_{\rm L,1}$ &2.466$_{-0.018}^{+0.017}$ &  2.590$_{-0.026}^{+0.026}$ &2.578$_{-0.012}^{+0.006}$   &2.838$_{-0.028}^{+0.029}$ & 3.116$_{-0.021}^{+0.022}$ &3.242$_{-0.017}^{+0.018}$ & 3.822$_{-0.029}^{+0.029}$ &  4.08$_{-0.05}^{+0.05}$ & 4.15$_{-0.07}^{+0.06}$ & 4.493$_{-0.024}^{+0.024}$  \\
$c_{\rm L,1}$  &0.298$_{-0.027}^{+0.03}$ & 0.32$_{-0.03}^{+0.04}$ & 0.41$_{-0.04}^{+0.04}$   &0.41$_{-0.05}^{+0.05}$  & 0.43$_{-0.04}^{+0.04}$ &0.336$_{-0.026}^{+0.027}$ & 0.45$_{-0.04}^{+0.03}$ & 0.482$_{-0.028}^{+0.013}$ &  0.458$_{-0.05}^{+0.029}$ & 0.466$_{-0.03}^{+0.022}$  \\
$a_{\rm L,2}$ &0.0058$_{-0.0010}^{+0.0011}$ &  0.0039$_{-0.0011}^{+0.0011}$ &0.0030$_{-0.0008}^{+0.0011}$   &0.0050$_{-0.0014}^{+0.0018}$  & 0.0036$_{-0.0008}^{+0.0010}$ &0.0037$_{-0.0005}^{+0.0005}$ & 0.0025$_{-0.0004}^{+0.0005}$ & 0.0018$_{-0.0006}^{+0.0005}$ & 0.00118$_{-0.0004}^{+0.0004}$ & 0.0012$_{-0.00023}^{+0.00024}$  \\
$b_{\rm L,2}$ &3.342$_{-0.04}^{+0.029}$ & 3.56$_{-0.05}^{+0.05}$ &  3.65$_{-0.05}^{+0.04}$  &3.91$_{-0.07}^{+0.06}$ & 4.18$_{-0.05}^{+0.05}$ &4.419$_{-0.029}^{+0.03}$ & 5.18$_{-0.05}^{+0.04}$ & 5.52$_{-0.09}^{+0.09}$ & 5.95$_{-0.09}^{+0.06}$ & 6.08$_{-0.07}^{+0.05}$  \\
$c_{\rm L,2}$ &0.17$_{-0.04}^{+0.04}$ & 0.16$_{-0.04}^{+0.07}$ &0.15$_{-0.04}^{+0.12}$  &0.30$_{-0.09}^{+0.11}$  & 0.30$_{-0.08}^{+0.09}$ &0.25$_{-0.04}^{+0.04}$ &  0.21$_{-0.05}^{+0.06}$ & 0.32$_{-0.10}^{+0.11}$ & 0.18$_{-0.06}^{+0.13}$ & 0.22$_{-0.05}^{+0.05}$   \\
$a_{\rm L,3}$&0.0313$_{-0.0011}^{+0.0011}$ & 0.0191$_{-0.0012}^{+0.0013}$ &0.0206$_{-0.0011}^{+0.0011}$  &0.0184$_{-0.0008}^{+0.0009}$ & 0.0161$_{-0.0006}^{+0.0006}$ &0.0119$_{-0.0004}^{+0.0004}$ & 0.0074$_{-0.0004}^{+0.0004}$ & 0.0068$_{-0.0006}^{+0.0006}$ & 0.0060$_{-0.0008}^{+0.0008}$ &  0.0038$_{-0.0003}^{+0.0003}$  \\
$b_{\rm L,3}$ &5.700$_{-0.009}^{+0.009}$ & 5.971$_{-0.017}^{+0.019}$ & 6.242$_{-0.016}^{+0.016}$  &6.629$_{-0.016}^{+0.016}$ & 7.115$_{-0.013}^{+0.014}$ &7.482$_{-0.012}^{+0.013}$ & 8.714$_{-0.021}^{+0.022}$ & 9.13$_{-0.03}^{+0.03}$ &   9.66$_{-0.05}^{+0.06}$ & 10.16$_{-0.05}^{+0.05}$ \\
$c_{\rm L,3}$&0.278$_{-0.013}^{+0.013}$ &  0.234$_{-0.022}^{+0.024}$ &0.366$_{-0.025}^{+0.028}$    &0.354$_{-0.026}^{+0.028}$ & 0.363$_{-0.018}^{+0.020}$ &0.346$_{-0.018}^{+0.018}$ & 0.346$_{-0.024}^{+0.029}$ &  0.31$_{-0.04}^{+0.04}$ & 0.39$_{-0.07}^{+0.07}$ & 0.460$_{-0.04}^{+0.029}$   \\
$a_{\rm L,4}$&0.0290$_{-0.0025}^{+0.0024}$ &  0.0297$_{-0.0015}^{+0.0016}$ &0.0175$_{-0.0010}^{+0.0010}$    &0.0206$_{-0.0007}^{+0.0007}$ & 0.0133$_{-0.0005}^{+0.0005}$ &0.011$_{-0.0004}^{+0.0004}$ &  0.0075$_{-0.0004}^{+0.0004}$ & 0.0057$_{-0.0006}^{+0.0006}$ & 0.0041$_{-0.0007}^{+0.0007}$ & 0.0025$_{-0.0003}^{+0.0003}$  \\
$a_{\rm L,4b}$\textdagger & 0.0104$_{-0.0016}^{+0.0019}$  & -& -& -& -& -& -& -& -& -\\
$b_{\rm L,4}$ &7.066$_{-0.028}^{+0.028}$ & 7.551$_{-0.026}^{+0.026}$ & 7.916$_{-0.029}^{+0.028}$   &8.413$_{-0.029}^{+0.027}$ & 9.058$_{-0.026}^{+0.025}$ &9.470$_{-0.029}^{+0.029}$ & 10.76$_{-0.05}^{+0.05}$ & 11.44$_{-0.06}^{+0.06}$ &12.56$_{-0.18}^{+0.15}$ & 12.56$_{-0.11}^{+0.13}$ \\
$b_{\rm L,4b}$\textdagger & 7.615$_{-0.020}^{+0.019}$   & -& -& -& -& -& -& -& -& -\\
$c_{\rm L,4}$ &\textdagger0.438$_{-0.03}^{+0.029}$ & 0.450$_{-0.03}^{+0.028}$ &0.460$_{-0.027}^{+0.026}$  &0.4965$_{-0.006}^{+0.0026}$ & 0.493$_{-0.011}^{+0.005}$ &0.4970$_{-0.005}^{+0.0021}$ &  0.494$_{-0.011}^{+0.005}$ & 0.462$_{-0.05}^{+0.025}$ & 0.466$_{-0.05}^{+0.027}$ &  0.485$_{-0.025}^{+0.012}$   \\
$c_{\rm L,4b}$\textdagger & 0.215$_{-0.025}^{0.028}$ & -& -& -& -& -& -& -& -& -\\
$a_{\rm L,5}$&0.0031$_{-0.0009}^{+0.0009}$ & 0.0025$_{-0.0008}^{+0.0008}$ &0.0031$_{-0.0007}^{+0.0007}$  &0.0018$_{-0.0004}^{+0.0004}$ & 0.0009$_{-0.0003}^{+0.0003}$ &0.00072$_{-0.00021}^{+0.00022}$ &            -               &            -               &            -               &            -               \\          
$b_{\rm L,5}$ &11.74$_{-0.17}^{+0.18}$& 12.32$_{-0.15}^{+0.18}$ &13.11$_{-0.14}^{+0.4}$ &13.47$_{-0.13}^{+0.12}$& 14.39$_{-0.08}^{+0.07}$ &14.92$_{-0.08}^{+0.11}$ &            -               &            -               &            -               &            -               \\         
$c_{\rm L,5}$ &0.86$_{-0.20}^{+0.10}$ & 0.42$_{-0.14}^{+0.07}$ &0.46$_{-0.09}^{+0.04}$   &0.42$_{-0.10}^{+0.06}$ & 0.28$_{-0.10}^{+0.12}$  &0.25$_{-0.08}^{+0.10}$ &            -               &            -               &            -               &            -               \\         
\hline 
\end{tabular}
\begin{tablenotes} 
\item \textdagger In the first dataset the position around 7-8\arcmin was fitted with two Lorentzian functions. The parameters of the second Lorentzian are indicated with the subscript 'b'.
\item \textit{Notes --} Lorentzians are defined in  Eq.~(\ref{eq:lor}) with $a_{\rm L}$ being the normalization (in units of counts s$^{-1}$ arcmin$^{-1}$), $b_{\rm L}$ being the center in units of arcmin and  $c_{\rm L}$ being the half-width at half-maximum also in units of arcmin.
\end{tablenotes}
\end{threeparttable}
\end{adjustbox}
\end{table*}

\end{appendix}



\bsp	
\label{lastpage}
\end{document}